%% file: ms.tex
\documentclass[useAMS,usenatbib]{mn2e}
\usepackage{graphics, graphicx}
\newcommand{\NrHighEdotGlast}{61}
\newcommand{\TotNrPulsars}{352}
\newcommand{\SlopeWPrelation}{$-0.30\pm0.05$}
\newcommand{\SlopeWPrelationwithouterror}{$-0.30$}

\newif\ifAMStwofonts

\def\lapp{\ifmmode\stackrel{<}{_{\sim}}\else$\stackrel{<}{_{\sim}}$\fi}
\def\gapp{\ifmmode\stackrel{>}{_{\sim}}\else$\stackrel{>}{_{\sim}}$\fi}

\newcommand{\degrees}[1]{\ensuremath{#1^\circ}}


\title[Characteristics of energetic pulsars]
{Profile and polarization characteristics of energetic pulsars}
\author[Weltevrede \& Johnston]
{Patrick Weltevrede and Simon Johnston \\
Australia Telescope National Facility, CSIRO, P.O. Box 76, 
Epping, NSW 1710, Australia.
}
\date{\today}
\pagerange{\pageref{firstpage}--\pageref{lastpage}}
\pubyear{2008}
\begin{document}
\maketitle
\label{firstpage}

\begin{abstract}
In this paper we compare the characteristics of pulsars with a high
spin-down energy loss rate ($\dot{E}$) against those with a low
$\dot{E}$.  We show that the differences in the total intensity pulse
morphology between the two classes are in general rather subtle.  A
much more significant difference is the fractional polarization which
is very high for high $\dot{E}$ pulsars and low for low $\dot{E}$
pulsars.  The $\dot{E}$ at the transition is very similar to the death
line predicted for curvature radiation. This suggests a possible link
between high energy and radio emission in pulsars and could imply that
$\gamma$-ray efficiency is correlated with the degree of linear
polarization in the radio band.  The degree of circular polarization
is in general higher in the second component of doubles, which is
possibly caused by the effect of co-rotation on the curvature of the
field lines in the inertial observer frame.

The most direct link between the high energy emission and the radio
emission could be the sub-group of pulsars which we call the energetic
wide beam pulsars. These young pulsars have very wide profiles with
steep edges and are likely to be emitted from a single magnetic
pole. The similarities with the high energy profiles suggest that both
types of emission are produced at the same extended height range in
the magnetosphere.  Alternatively, the beams of the energetic wide
beam pulsars could be magnified by propagation effects in the
magnetosphere. This would naturally lead to decoupling of the wave
modes, which could explain the high degree of linear polarization. As
part of this study, we have discovered three previous unknown
interpulse pulsars (and we detected one for the first time at
20~cm). We also obtained rotation measures for 18 pulsars whose values
had not previously been measured.

\end{abstract}


\section{Introduction}

Pulsars are observed to be spinning down with time.  The spin-down
energy loss rate $\dot{E}$, which is the loss of kinetic energy, is
given by
\begin{equation}
\dot{E} = 4\pi^2I\dot{P}P^{-3}
\end{equation}
where $I$ is the moment of inertia of the star (generally taken to be
$10^{45}$~g$\,$cm$^{2}$), $P$ its spin period and $\dot{P}$ its spin
down rate. Some of the loss of spin-down energy emerges as radiation
across the entire electromagnetic spectrum from radio to
$\gamma$-rays. The radio emission accounts for only $\sim10^{-6}$ of the
energy budget (e.g. \citealt{lk05}) whereas up to a few percent is
emitted in the $\gamma$-ray band (e.g. \citealt{tho04}), with the rest
converted to magnetic dipole radiation and some form of pulsar wind.

It has been evident for more than a decade that pulsars with high
$\dot{E}$ have different polarization characteristics to those with
lower $\dot{E}$. Many high $\dot{E}$ pulsars are highly linearly
polarized (e.g. \citealt{qmlg95,hlk98,cmk01}). The pulse profiles of
high $\dot{E}$ pulsars are believed to be generally simple, consisting
of either one or two prominent components
(e.g. \citealt{hmt71,ran83}). \cite{jw06} found that, in the high
$\dot{E}$ pulsars with double profiles, the total power and the
circular polarization usually dominates in the trailing component and
that the swing of position angle (PA) of the linearly polarized
radiation is steeper under the trailing component. They interpreted
these results as showing that the beam of high $\dot{E}$ pulsars
consisted of a single conal ring at a relatively high
height. \cite{kj07} incorporated these results into their pulsar beam
model. In their model, there is a sharp distinction between pulsars
with $\dot{E}>10^{35}$ erg~s$^{-1}$ and those with smaller $\dot{E}$.

High $\dot{E}$ pulsars are not only interesting because of their
distinct properties in the radio band, but also because a subset of
them emit pulsed high energy emission. There are three different
families of high energy emission models in the literature which places
the emitting regions at different locations in the pulsar
magnetosphere. In the polar cap models (e.g. \citealt{dh96}) the
emitting region is close to the neutron star surface, while outer gap
models (e.g. \citealt{chr86a}) place the emitting region near the
light cylinder. Finally, in slot gap models (e.g. \citealt{mh04}) the
particle acceleration occurs in a region bordering the last open field
lines. In the polar cap models the young pulsars are thought to
produce pairs through curvature radiation (e.g. \citealt{hm01a}),
while older pulsars produce pairs only through inverse-Compton
scattering (e.g. \citealt{mh04b}). In the outer gap model pairs are
formed by the interaction of thermal X-ray photons from the neutron
star surface with $\gamma$-ray photons (e.g. \citealt{rom96a}). All
models have in common that high $\dot{E}$ pulsars should be brighter
$\gamma$-ray sources than low $\dot{E}$ pulsars, something which
is confirmed by EGRET (\citealt{tho04}).

We have recently embarked on a long-term timing campaign to monitor a
large sample of young, high $\dot{E}$ pulsars. The ephemerides
obtained from timing will be used to provide accurate phase tagging of
$\gamma$-ray photons obtained from the Fermi Gamma-Ray Space Telescope
(formerly known as GLAST \citealt{sgc+08}) and AGILE
(\citealt{ppp+08}) satellites, with the expectation that the number of
$\gamma$-ray pulsars will increase from the current 7 to over 100
(\citealt{gsc+07}). Of the $\sim80$ non millisecond pulsars in the
pulsar catalogue maintained by the
ATNF\footnote{http://www.atnf.csiro.au/research/pulsar/psrcat}
\citep{mhth05} with $\dot{E}>10^{35}$ erg~s$^{-1}$, we have obtained
polarization profiles at 1.4~GHz for \NrHighEdotGlast, a substantial
increase in the number available to previous studies. In this paper,
therefore, we examine the differences between high and low $\dot{E}$
pulsars. In total, we use pulse profiles from {\TotNrPulsars} pulsars,
which includes the {\NrHighEdotGlast} energetic pulsars and a
comparison sample of intermediate and low $\dot{E}$ pulsars in order
to draw general conclusions about the pulsar population.

The paper is organized as follows. We start with explaining the
details of the observations and the data analysis. In section 3 we
then describe the polarization profiles of four pulsars for which we
found an interpulse at 20~cm and present new rotation measures. In
section 4 the the total intensity profiles of the pulsars are
discussed, followed by a discussion of the polarization
properties. Finally we will discuss the results in section 6, followed
by the conclusions. The polarization profiles of all the pulsars can
be found in appendix A (those for which we have a 20~cm and a 10~cm
profile) and appendix B (those for which we only have a 20~cm
profile). The plots of the pulse profiles can also be found on the
internet\footnote{http://www.atnf.csiro.au/people/joh414/ppdata/index.html}.
Finally, a table with derived properties from the pulse profiles can
be found in appendix C\footnote{This table is also available in
electronic form at the CDS via anonymous ftp to {\tt
cdsarc.u-strasbg.fr (130.79.128.5)} or via {\tt
http://cdsweb.u-strasbg.fr/cgi-bin/qcat?J/MNRAS/}}. The appendices are
only available in the on-line version of this publication.

\section{Observations and data analysis}

The procedure to generate pulse profiles for the pulsars which are
timed for the Fermi and AGILE satellites is complicated by the
fact that the pulse profiles of individual (short) observations have
typically a low signal to noise ratio ($S/N$). It is therefore
required to sum all the available observations in order to obtain a
template profile with a higher $S/N$. This procedure is described in
some detail in this section.

\subsection{Observations}

All the observations were made at the Parkes telescope in
Australia using the centre beam of the 20~cm multibeam receiver (which
has a bandwidth of 256~MHz and has a noise equivalent flux density of
$\sim$35 Jy on a cold sky) and the 10/50~cm receiver (which has at
10~cm a bandwidth of 1024~MHz and has a noise equivalent flux density of
$\sim$49 Jy on a cold sky). This paper will focus mainly on the 20~cm
data, because that is wavelength at which the majority of observations
were made. However for some highly scattered pulsars it is also useful
to consider the 10~cm data. The 50~cm data is not used, because the
profiles are scattered at that frequency in many cases. The timing
program started in April 2007 and each pulsar is typically observed
once per month at 20~cm and twice per year at 10 and 50~cm. The two
polarization channels of the linear feeds of the receiver were
converted into Stokes parameters, resampled and folded at the pulse
period by a digital filterbank. In our case a pulse profile with 1024
bins and 1024 frequency channels was dumped every 30 seconds on hard
disk. Before each observation a calibration signal, injected into the
feed at a \degrees{45} angle to the probes, was recorded which is then
used to determine the phase delay and relative gain between the two
polarization channels.

The data were processed using the {\sl PSRCHIVE} package
(\citealt{hvm04}). The data of each observing session were first
checked for narrow band radio frequency interference (RFI). An
automatic procedure using the median smoothed difference of the
bandpass, was used to identify the affected frequency channels in the
calibration observations. The flagged channels were left out of all
the observations of a particular observing day, making the automatic
procedure more robust in finding weaker RFI which is not always
identified. The remaining frequency channels were added together and
the resulting sequence of profiles was then visually inspected for
impulsive RFI. The sub-integrations in where RFI was particular strong
were left out of further data processing.

The 20~cm multibeam receiver has a significant cross-coupling between
the two dipoles affecting the polarization of the pulsar signal. For
instance, a highly linearly polarized signal induces artificial
circular polarization. These effects are measured as a function of
parallactic angle for PSR J0437--4715 for the Parkes Pulsar Timing
Array project (Manchester et al. 2008, in preparation), which allows
the construction of a polarimetric calibration model
(\citealt{van04}). We have applied this model to all the observations
using the 20~cm multibeam receiver, which reduces the artifacts in the
Stokes parameters considerably.

\subsection{\label{SctSumming}Summing of the individual observations}

For some pulsars the timing noise is so severe that the pulse period
predicted by the timing solution in the pulsar catalogue is not
accurate enough to fold the data. In such a case the pulsar appears to
drift in longitude in successive sub-integrations. We therefore
applied the updated timing solutions to align the sub-integrations
within individual observations.

To produce high $S/N$ profiles the individual observations must be
added together. Because many pulsars involved in this timing program
have severe timing noise and show glitches, it is difficult to use the
timing solution to add the observations together. Instead a scheme was
followed in which the observations are correlated with each other in
order to find the offsets in pulse longitude between the
profiles. These offsets were applied directly to the individual
observations using custom software in order to sum the profiles. The
sum of the profiles (i.e. the standard or template) has a higher $S/N$
than the individual profiles and can then be correlated with the
individual observations to determine the offsets in pulse longitude
with higher precision, hence making a more accurate standard. This
procedure is repeated one more time to make the final pulse profile.

\subsection{Faraday de-rotation}

The interstellar medium interacts with the radio waves of pulsars,
causing a number of frequency and time dependent effects. One of these
effects is Faraday rotation, where the interstellar magnetic field
component parallel to the line of sight causes a difference in the
propagation speeds of the left- and right-hand circular polarization
signal components. This effect causes the polarization vector to
rotate in the Stokes Q and U plane and the angle is a function of
frequency and the rotation measure (RM). It is therefore necessary to
de-rotate Stokes Q and U before summing the frequency channels in an
observation.

A similar procedure has to be followed when the profiles of different
observations are summed together, because different frequency channels
were flagged and deleted in different observations. This means that
although the centre frequencies are identical for the different
observations, their weighted mid-frequencies are slightly
different. The {\sl PSRCHIVE} package de-rotates Stokes Q and U with
respect to this weighted mid-frequency of the band and therefore it is
necessary to take the RM into account when profiles of different
observations are summed together. This is done by rotating Stokes Q
and U of each observation with respect to infinite frequency using
custom software before adding individual observations together.

\subsection{Making frequency standards}

In order to be able to measure the RM for pulsars for which no
sufficiently accurate values were available one needs to keep
frequency resolution. This is done by summing the observations
together using custom software which takes into account the pulse
longitude offsets found by correlating the profiles of the individual
observations (as described section \ref{SctSumming}). A complication
is that {\sl PSRCHIVE} de-disperses the data with respect to the
non-weighted centre frequency of the band, while the pulse longitude
offsets are determined using de-dispersed profiles with respect to the
weighted mid-frequency. It is therefore necessary to include a
dispersion time delay corresponding with the difference in the
weighted and non-weighted mid-frequency when the observations are
added together.

\section{Results on individual pulsars}

\subsection{Newly discovered interpulses}

\begin{figure}
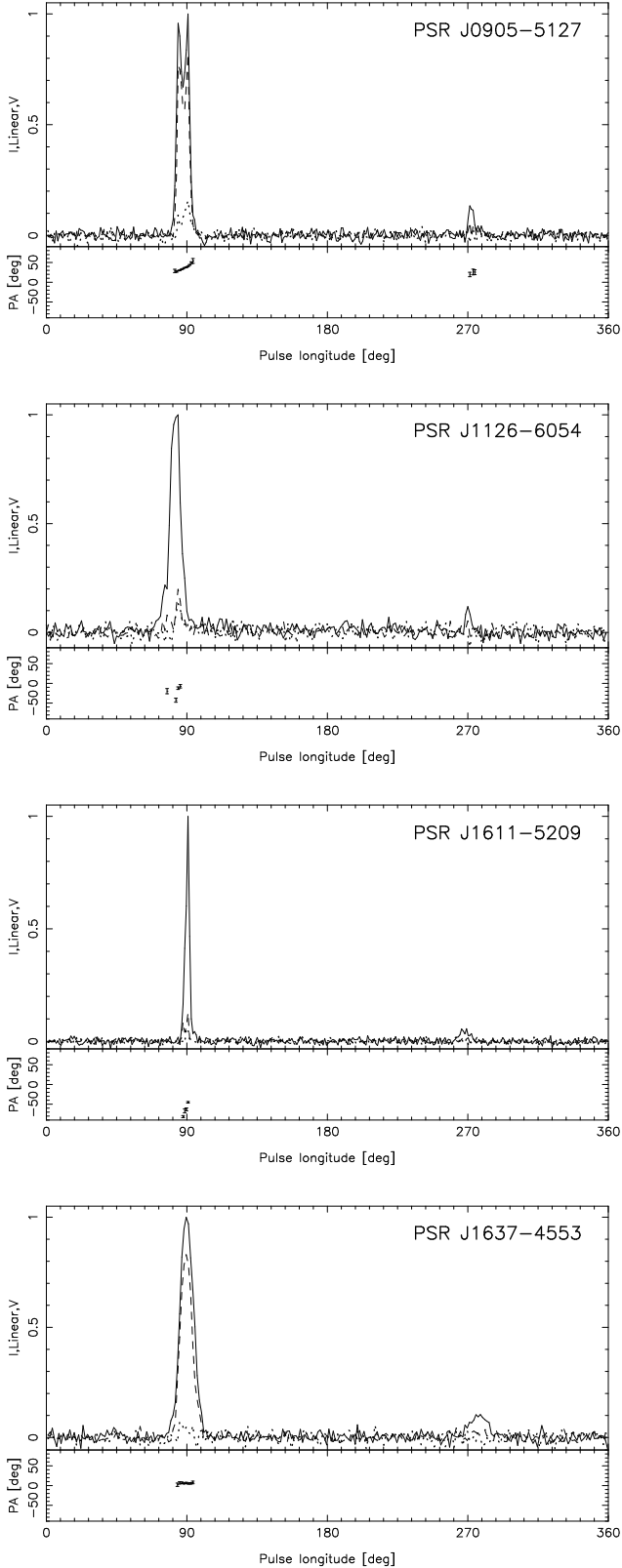

\begin{center}
\includegraphics[height=0.99\hsize,angle=270]{J0905-5127.20.paperplot.ps}\\
\vspace*{5mm}
\includegraphics[height=0.99\hsize,angle=270]{J1126-6054.20.paperplot.ps}\\
\vspace*{5mm}
\includegraphics[height=0.99\hsize,angle=270]{J1611-5209.20.paperplot.ps}\\
\vspace*{5mm}
\includegraphics[height=0.99\hsize,angle=270]{J1637-4553.20.paperplot.ps}
\caption{\label{Fig_newIP}The pulse profile of the four pulsars for
which we report an interpulse at an observing wavelength of 20~cm.
The top panels show the total intensity profile (solid line), linear
polarization (dashed line) and circular polarization (dotted
line). The peak intensity of the profiles are normalized to one. The bottom
panels show the PA of the linear polarization (for the pulse longitude
bins in where the linear polarization was detected above $2 \sigma$).}
\end{center}
\end{figure}

While analysing the data described in this paper we discovered four
interpulses which have not been previously reported at 20~cm in the
literature. The polarization profiles of these pulsars are shown in
Fig. \ref{Fig_newIP} and discussed in some detail below. In all
cases there is no evidence that the interpulse only appears
sporadically rather than be weakly present in all observations.

\subsubsection{PSR J0905--5127}

Profiles of this pulsar were presented first in \cite{dsb+98}. In
their figure, there appears to be little sign of the interpulse at
20~cm, with perhaps a hint at 70~cm.  In our observations at 20~cm,
the interpulse is very weak in comparison to the main pulse with an
intensity ratio of $\sim$17. The separation between the centroid of
the main and interpulses is 175\degr. The main pulse is a clear double
with a total width of $\sim\degrees{20}$ but not much structure can be
discerned in the interpulse because of the low $S/N$, although its
width appears to be narrower than that of the main pulse. The
interpulse is separated by \degrees{180} from the trailing component
of the main pulse, suggesting that the interpulse could be the
trailing component of a double. The polarization swing across both the
main and interpulses is rather flat, and we cannot attempt a rotating
vector model fit (RVM; \citealt{rc69a}). We do not have sufficient
$S/N$ to make any claims about the interpulse at either 10~cm or 50~cm
from our data.

\subsubsection{PSR J1126--6054}

The interpulse of PSR J1126--6054 is too weak to be seen in the
profile presented in \cite{jlm+92}. However, it is just visible in our
20~cm data with a peak amplitude about one tenth of that of the main
pulse. The peak-to-peak separation between the main and interpulses is
$\sim$174\degr. This low $\dot{E}$ pulsar has a low degree of linear
polarization in its main pulse, which explains the absence of
significant linear polarization in the much weaker interpulse.
The interpulse appears to be significantly narrower, though the low
$S/N$ makes the width difficult to measure. At 50~cm, the interpulse
is marginally stronger with respect to the main pulse whereas at 10~cm
it is not detected.

\subsubsection{PSR J1611--5209}

There is no obvious interpulse at 20~cm in the profile presented in
\cite{jlm+92}. However, \cite{kjm05} reported a low-amplitude
interpulse in their 10~cm data.  In our 20~cm data we clearly see the
interpulse which has a peak amplitude less than 0.1 that of the main
pulse. The separation between the main and interpulse is
$\sim$177\degr. The main pulse has a total width of $\sim$10\degr\ and
consists of at least two components with a low fractional
polarization. The low $S/N$ in the interpulse precludes any
measurement of the polarization, but the overall width seems similar
to main pulse.

\subsubsection{PSR J1637--4553}

This pulsar has a very weak interpulse (about one tenth in amplitude
compared to the main pulse), which is perhaps just visible in the
existing literature (\citealt{jlm+92}). The separation between the
main and interpulses is $\sim\degrees{173}$, and although weak, the
interpulse seems to be the same width as the $\sim\degrees{20}$ of the
main pulse. The polarization of the interpulse is hard to determine,
although the main pulse is virtually 100\% polarized. At 50~cm, the
interpulse has the same separation from the main pulse and roughly the
same relative amplitude as at 20~cm. Our low $S/N$ at 10~cm makes the
interpulse undetectable.

\subsection{New rotation measures}

\begin{table}
\begin{tabular}{crrrlr}
\hline
Name & \multicolumn{1}{c}{RM} & \multicolumn{1}{c}{$l$} & \multicolumn{1}{c}{$b$} & \multicolumn{1}{c}{DM} & \multicolumn{1}{c}{$d$}\\
 & \multicolumn{1}{c}{[rad m$^{-2}$]} & \multicolumn{1}{c}{[deg]} & \multicolumn{1}{c}{[deg]} & \multicolumn{1}{c}{[pc cm$^{-3}$]} & \multicolumn{1}{c}{[kpc]}\\
\hline
J1052--5954 & $-280\pm24$      & 288.55 & -0.40 &  \hspace*{3mm}491    &   13.55  \\
J1115--6052 & $257\pm18$       & 291.56 & -0.13 &  \hspace*{3mm}228.2  &   6.76   \\
J1156--5707 & $238\pm19$       & 295.45 & 4.95  &  \hspace*{3mm}243.5  &   20.40  \\
J1524--5625 & $180\pm20$       & 323.00 & 0.35  &  \hspace*{3mm}152.7  &   3.84   \\
J1524--5706 & $-470\pm20$      & 322.57 & -0.19 &  \hspace*{3mm}833    &   21.59  \\
J1638--4417 & $160\pm25$       & 339.77 & 1.73  &  \hspace*{3mm}436.0  &   8.46   \\
J1702--4128 & $-160\pm20$      & 344.74 & 0.12  &  \hspace*{3mm}367.1  &   5.18   \\
J1705--3950 & $-106\pm14$      & 346.34 & 0.72  &  \hspace*{3mm}207.1  &   3.86   \\
J1737--3137 & $448\pm17$       & 356.74 & 0.15  &  \hspace*{3mm}488.2  &   5.88   \\
J1738--2955 & $-200\pm20$      & 358.38 & 0.72  &  \hspace*{3mm}223.4  &   3.91   \\
J1801--2154 & $160\pm40$       & 7.83   & 0.55  &  \hspace*{3mm}387.9  &   5.15   \\
J1809--1917 & $41\pm17$        & 11.09  & 0.08  &  \hspace*{3mm}197.1  &   3.71   \\
J1815--1738 & $175\pm20$       & 13.18  & -0.27 &  \hspace*{3mm}728    &   9.06   \\
J1828--1101 & $45\pm20$        & 20.49  & 0.04  &  \hspace*{3mm}607.4  &   7.26   \\
J1837--0604 & $450\pm25$       & 25.96  & 0.26  &  \hspace*{3mm}462    &   6.19   \\
J1841--0345 & $447\pm15$       & 28.42  & 0.44  &  \hspace*{3mm}194.32 &   4.15   \\
J1845--0743 & $440\pm12$       & 25.43  & -2.30 &  \hspace*{3mm}281.0  &   5.85   \\
J1853--0004 & $647\pm16$       & 33.09  & -0.47 &  \hspace*{3mm}438.2  &   6.58   \\
\hline
\end{tabular}
\caption{\label{newRMs}The pulsars for which new values of the RM were
measured. From left to right the columns are the pulsar name, the
measured rotation measures, the galactic longitude and latitude, the
dispersion measure and the best available distance estimate.}
\end{table}

As mentioned in section 2, the interstellar magnetic field parallel to
the line of sight causes the polarization vector to rotate in the
Stokes Q and U plane. In order to derive the degree of linear
polarization it is therefore necessary to correct for this rotation
before summing the frequency channels across the frequency band. The
amount of rotation of the PA depend on the rotation measure RM and
values for the RM were obtained from the pulsar catalogue. However,
not all the pulsars have a published value for its RM (or one with
sufficient accuracy). We therefore measured the RM for a number of
objects in our sample.

The RM can be measured by fitting the change of the PA ($\psi$)
across the frequency band with the Faraday rotation formula
\begin{equation}
\label{EqRM}
\psi\left(\lambda\right)=\psi_\infty+RM\,\lambda^2,
\end{equation}
where $\lambda$ is the observing wavelength of the considered
frequency channel and $\psi_\infty$ is the PA at infinite
frequency. When different pulse longitude bins of the pulse profile
show a similar frequency dependence of $\psi$ one can be confident in
the measured RM. The RM is obtained by calculating the weighted
average of the fits of equation (\ref{EqRM}) for different bins in
where there is enough linear polarization present. The new RM values
are listed in Table \ref{newRMs}. Only PSR J1809--1917 has a
previously published RM which we include in the table because our
value of 41 rad m$^{-2}$ differs significantly from the 130 rad
m$^{-2}$ quoted by Han et al. (2006).

\section{Total intensity pulse profiles}

In this and the following section we investigate if, and how, the
beams of high $\dot{E}$ pulsars differ from those of low $\dot{E}$
pulsars. As we are going to investigate basic pulse profile properties
in a statistical way, it is important to consider the effects of a low
$S/N$ and interstellar scattering. Because the high $\dot{E}$ pulsars
tend to be younger they have on average lower galactic latitude than
older pulsars, hence they tend to be more affected by interstellar
scattering. Therefore low $S/N$ observations and profiles which are
clearly affected by interstellar scattering were excluded from the
statistics.

\subsection{Pulse profile morphology}

It has been pointed out by several authors
(e.g. \citealt{hmt71,ran83,jw06,kj07}) that the profiles of high
$\dot{E}$ pulsars are relatively simple. A problem with measuring
``profile complexity'' is that it is not a well defined quantity,
hence it is highly subjective. In order to make the results objective
and better reproducible one should quantify the amount of complexity
in a mathematical way. We will therefore explore ways to quantify
different aspects of profile complexity, because it is difficult to
come up with a definition which covers all facets of profile
complexity.
Only the total intensity (Stokes parameter I) profiles are
considered in this section, while the polarization properties are
investigated in the next section.

\subsubsection{Profile classification}

\begin{table}
\begin{tabular}{c@{\,} c@{} c c c c c}
\hline
\multicolumn{3}{c}{$\dot{E}$ [erg~s$^{-1}$]} & Single & Double & Multiple & Total\\
\hline
$10^{35}$ & -- & $10^{38}$ & 27 (53\%) & 17 (33\%) & 7 (14\%) & 51\\
$10^{33}$ & -- & $10^{35}$ & 53 (47\%) & 43 (38\%) & 16 (14\%) & 112\\
$10^{28}$ & -- & $10^{33}$ & 52 (46\%) & 46 (40\%) & 16 (14\%) & 114\\
\hline
\end{tabular}
\caption{\label{classification}The classification of the profiles for
different $\dot{E}$ bins. Pulsars with a $S/N < 30$ were excluded as
well as the profiles marked to show substantial scattering.}
\end{table}

Pulse profiles are often described in terms of ``components'', which
are attributed to structure in the pulsar beam. There are different
models in the literature describing the structure of the radio beam of
pulsars. The beam could be composed out of a core and one or more
cones (\citealt{ran83}), randomly distributed patches (\citealt{lm88})
or patchy cones (\citealt{kj07}). In these models each component of
the pulse profile originates from a different physical location in the
magnetosphere. Because the components overlap in many cases and
because their shapes are not uniquely defined, it is difficult to
objectively classify profiles. Following \cite{kj07}, we have
classified the profiles by eye into three classes depending on the
number of distinct (possibly overlapping) peaks in the pulse
profile. These classes are named ``single'', ``double'' and
``multiple'', depending on if one, two or more peaks were
identified. Although this classification is subjective, it should be
considered as a rough measure for the complexity.

Table \ref{classification} shows the percentage of pulsars in each
class for three different $\dot{E}$ bins.  For the pulsars which are
significantly scattered at 20~cm we used 10~cm data (when available)
and we omitted profiles with $S/N<30$ to improve
significance. Compared with \cite{kj07} we find relatively more
singles and less multiples and also the difference between high and
low $\dot{E}$ pulsars is less pronounced. This might partially reflect
the subjectivity of profile classification, but it may also be related
to the fact that the classification of \cite{kj07} was based on
polarization properties. For instance, rapid changes in the PA-swing
are often found in between components and can therefore be interpreted
as an indication for the presence of multiple components.  The
polarization properties are discussed in a separate section in this
paper.

\subsubsection{\label{SctMathDecomp}Mathematical decomposition of the profiles}

Because profile components can overlap and can have various shapes, it
is in many cases not clear how many separate emission components there
are. Also, because the classification is done by eye it is highly
subjective at which level of detail the profile is separated into
components. A more objective way to decompose the profile into
components is to describe the profiles as linear combinations of basis
functions. The number of required functions to fit the profile
is then a measure for the complexity of the pulse profiles.

\begin{figure}
\begin{center}
\includegraphics[height=0.99\hsize,angle=270]{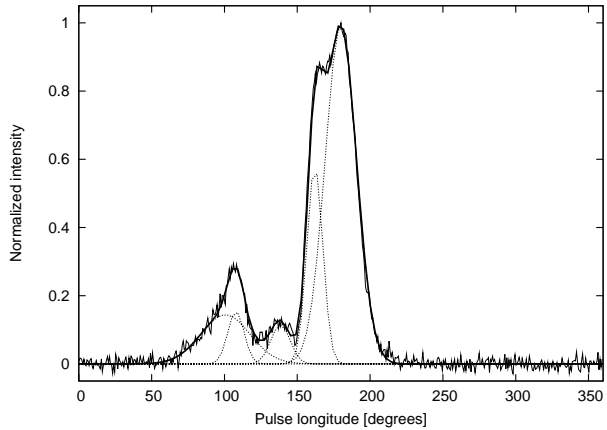}
\caption{\label{Fig_vonmises}The decomposition of the pulse profile of
PSR J1803--2137 at 20~cm into five von Mises functions (the dotted
curves). The sum of these functions (the thick solid line) is a good
representation of the observed pulse profile (thin solid line).}
\end{center}
\end{figure}

Gaussian functions are often used to decompose profiles
(e.g. \citealt{kwj+94}), but we have chosen to use von Mises functions
(\citealt{von18}), which are defined as
\begin{equation}
\label{Eqvonmises}
I(x) = I_\mathrm{peak}e^{\kappa\cos\left(x-\mu\right) - \kappa}.
\end{equation}
Here $\mu$ is the location of the peak (in radians), $I_\mathrm{peak}$
is the peak intensity and $\kappa$ is the concentration (which
determines the width of the peak). The shape of these functions is
very similar to Gaussians (see Fig. \ref{Fig_vonmises}), but they can
often fit the edges of components slightly better. The main difference
is that von Mises functions are circular, hence they are also known as
circular normal distributions. A fitting routine for von Mises
functions is part of the {\sl PSRCHIVE} software package. 

There is a subtle difference between the required number of fit
functions and the number of components in the pulse profile. The first
is just a mathematical measure of complexity, while the latter is the
number of distinct physical emission locations in the pulsar
magnetosphere which are visible in the line of sight. These numbers
can be different, because there is no a priori reason to believe that
the shape of a profile component can be described by a single, simple,
mathematical basis function which is the same for all pulsars.  For
instance, a profile which shows a tail because of interstellar
scattering can have one component (``single profile''), but it can
only be fitted by a number of von Mises functions. Another example can
be seen in the decomposition as shown in
Fig. \ref{Fig_vonmises}. Although the component between pulse
longitude \degrees{70} and \degrees{120} is fit by two von Mises
functions, the smooth shape does suggest that it is a single
asymmetric emission component. By using more complex asymmetric
mathematical functions it might be possible to decompose some profiles
in a smaller number of fit functions. However, in effect this is the
same as to fit a larger number overlapping more simple symmetric
functions which have less fit-parameters per function. 

There is not always one unique solution for the decomposition of a
profile and therefore the decomposition does not necessarily give
additional insight in how profiles are composed out of distinct
physical components. Nevertheless a noise free mathematical
description of a pulse profiles can be used as a measure for its
complexity.  Moreover it is a very useful technique which makes it
easier to measure profile properties such as pulse widths. An
additional advantage of a mathematical description of the profile is
that one can more accurately determine the component widths for
pulsars which have overlapping components.

\begin{figure}
\begin{center}
\includegraphics[height=0.95\hsize,angle=270]{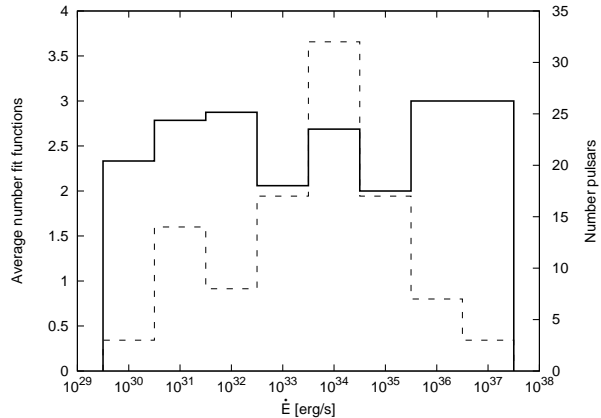}
\caption{\label{Fig_ppdot_complexity}The histogram of the average
number of von Mises functions required to fit the profiles when the
$S/N$ is scaled down to 30 for different $\dot{E}$ bins (solid
line). The dashed histogram shows the number of pulsars contributing
in each bin. Only profiles with a $S/N\geq100$ are included.
}
\end{center}
\end{figure}

When using the number of mathematical fit functions as a measure
of complexity it is important to take into account the $S/N$ ratio
of the profiles. A higher $S/N$ profile will require a larger number
of mathematical basis functions to fit its shape, even though the
profile is not necessarily more complex. In order to avoid this
effect, we determined how many of the fit functions would have a
significant contribution to the total integrated intensity of the
pulse profile when the $S/N$ would have been 30.  We only considered
profiles with a $S/N\ge100$ to ensure that all weak components which
are just significant when the $S/N$ would have been 30 are spotted by
eye.

Fig. \ref{Fig_ppdot_complexity} shows the average number of von Mises
functions required to fit the profiles when the $S/N$ is scaled down
to 30 for different $\dot{E}$ bins.  There is not much evidence that
the profile complexity is very different for high and low $\dot{E}$
pulsars. The absence of a significant correlation between $\dot{E}$
and the number of fit functions is confirmed by calculating the
Spearman rank-order correlation coefficient (\citealt{ptvf92}) of the
unbinned data, which is a non-parametric measure of correlation. Among
the most complex profiles, according to this classification scheme,
are those of PSRs J1034--3224 and J1745--3040, which indeed have
complex looking profile shapes.  An other pulsar which is ranked at
the same level of complexity is PSR J1302--6350, which has to the eye
a relatively simple double peaked profile, but its highly asymmetric
components require relatively many mathematical functions to fit.  We
will therefore try a different method to define profile complexity
below.

\subsubsection{The dimensionless double separation}

\begin{figure}
\begin{center}
\includegraphics[height=0.95\hsize,angle=270]{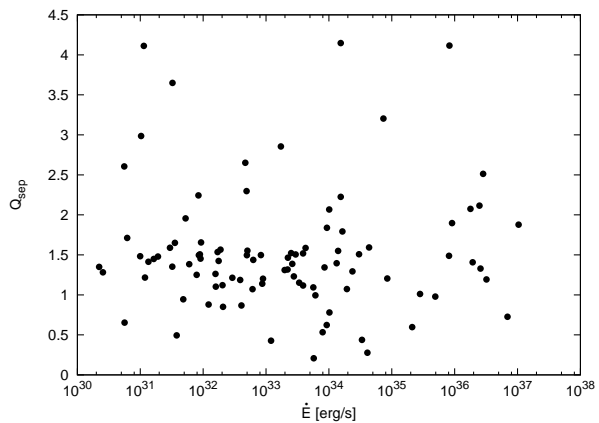}
\caption{\label{Fig_double_sep}The dimensionless double separation
(the ratio of the separation between the components and the average of
their full width half maxima) versus the spin-down energy loss rate
for all the observed pulsars at 20~cm which are classified to be
doubles with a $S/N\geq30$. }
\end{center}
\end{figure}

As described in section \ref{SctMathDecomp}, the profiles of the high
$\dot{E}$ pulsars J1015--5719 and J1302--6350 were ranked as highly
complex, while they appear to be ``simple'' to the eye. One property
of these profiles which makes them look simple is that they are
doubles with well separated components. We therefore tested the
hypothesis that the doubles of high $\dot{E}$ pulsars have more
clearly separated components than those of the low $\dot{E}$
pulsars. How clearly the components of doubles are separated can be
quantified by calculating a quality factor, which we define to be
\begin{equation}
Q_\mathrm{sep} = \frac{\mathrm{\Delta\phi_\mathrm{sep}}}{\frac{1}{2}\left(\mathrm{FWHM}_1+\mathrm{FWHM}_2\right)}.
\end{equation}
This dimensionless double separation is the ratio of the separation
between the components $\Delta\phi_\mathrm{sep}$ and the average of
the full width half maxima of the components $\mathrm{FWHM}_1$ and
$\mathrm{FWHM}_2$. Higher values of $Q_\mathrm{sep}$ imply that the
components are separated more compared with the width of the
components.

Fig. \ref{Fig_double_sep} shows $Q_\mathrm{sep}$ versus $\dot{E}$ for
all profiles at 20~cm which were classified to be doubles and have a
$S/N\geq30$. There is no evidence that the components of doubles of
low $\dot{E}$ pulsars are more likely to be overlapping, which is
confirmed by calculating the Spearman rank-order correlation
coefficient. According to this measure the most clearly separated
doubles are PSRs J1302--6350, J1733--3716, J1901--0906 and
J2346--0609.

\subsubsection{Profile symmetry}

A factor which was not taken into account in the previous sections is
the amount of symmetry in the profile. For instance, PSR J1302--6350
has highly asymmetric profile components, but the profile as a whole
appears symmetric and could therefore be regarded as ``simple''. It
is therefore interesting to consider the degree of symmetry of the
profiles, which can be
measured by cross-correlating the profile with its mirror-image. We
define the degree of profile symmetry to be the ratio of the maximum
value of the cross-correlation function between the profile and the
time reversed profile, and the maximum value of auto-correlation
function of the profile. The degree of symmetry is therefore
normalized to 1 for completely symmetric profiles and it decreases for
more asymmetric profiles.

\begin{figure}
\includegraphics[height=0.95\hsize,angle=270]{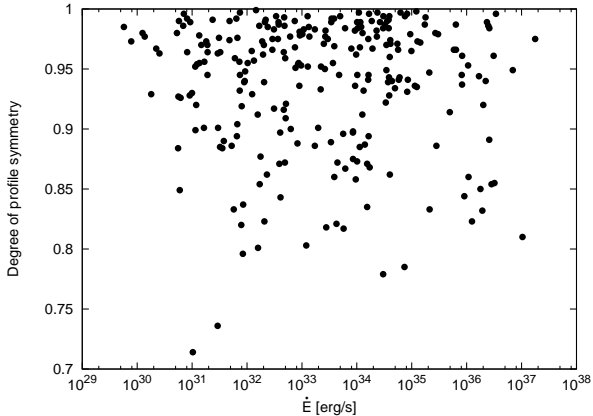}
\caption{\label{Fig_symmetry}The degree of profile symmetry versus
$\dot{E}$. Pulsars with a $S/N < 30$ and profiles with substantial
scattering were excluded and only 20~cm data was considered. There is
no evidence that high $\dot{E}$ pulsars are more symmetric.}
\end{figure}

The degree of symmetry versus $\dot{E}$ is shown in
Fig. \ref{Fig_symmetry}. The pulsar with the lowest measured degree of
symmetry is PSR B1747--31, which has a relatively narrow and bright
leading component and a much broader and weaker trailing
component. Also the complex main pulse of PSR B1055--52 can be found
at the lower end of this figure. There is no indication for any
correlation, which is confirmed by the Spearman rank-order correlation
coefficient. Like for the other measures of complexity, it is hard to
quantify that pulse profiles of high $\dot{E}$ pulsars are more simple
than those of the low $\dot{E}$ pulsars.

\subsection{\label{Sectbeamwidths}Pulse widths versus $P$}

A basic property of the emission beam of a pulsar is its half opening
angle $\rho$. It is found that the opening angle is proportional to
$P^{-1/2}$ (e.g. \citealt{big90b,kxl+98,gks93,ran93}),
which is expected if the edge of the active area of the polar cap is
set by the last open field lines. In order to derive the opening angle
from the measured profile width one needs to know how the emission
beam intersects the line of sight. Because the orientation of the line
of sight with respect to the pulsar beam is for most pulsars at best
only poorly constrained, it is difficult to obtain accurate opening
angles. For a large sample of pulsars the unknown geometrical factors
should average out and therefore the profile width and $\rho$ should
have the same $P$ dependence. The unknown geometry will cause
additional scattering around the correlation between the pulse width
and $P$.

\begin{figure}
\includegraphics[height=0.95\hsize,angle=270]{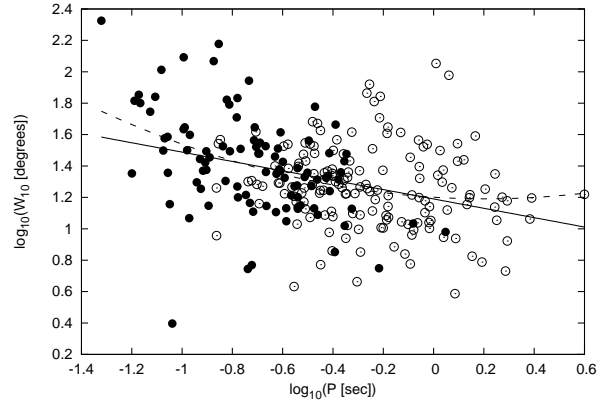}
\caption{\label{Fig_pulsewidths_p}The measured profile 10\% widths
versus $P$. The solid line is the power law fit through the data,
which has a slope of \SlopeWPrelationwithouterror. The dashed line
indices the fit of a second order polynomial through the data points
(which is statistically not better than the power law fit).  The
filled points are pulsars with an $\dot{E} \ge 10^{34}$ erg~s$^{-1}$
and the open points have a lower $\dot{E}$.  Pulsars with a $S/N < 30$
and profiles with substantial scattering were excluded. All the shown
observations were done at a wavelength of 20~cm.}
\end{figure}

The measured pulse widths at 10\% of the peak intensity ($W_{10}$)
indeed show a slight anti-correlation with $P$
(Fig. \ref{Fig_pulsewidths_p}), while there is no indication for a
dependence with $\dot{P}$ (not shown). The slope is measured by
reduced $\chi^2$ fitting (the data points are weighted equally), which
results in a slope of \SlopeWPrelation\footnote{The correlation
is slightly different, although within the error identical to the
value quoted in \cite{wj08a}. This is because since the publication of
that paper more timing observations have been added.}, comparable
with the fit obtained from the data of \cite{gl98} by
\cite{wj08a}. The slope of the correlation is therefore slightly less
than what is expected from theory. This conclusion, in combination
with the period distribution of pulsars with interpulses, provides
convincing evidence in favour for the evolution of the pulsar beam
towards alignment with the rotation axis (\citealt{wj08a}).

If there is any deviation from a power law relationship between
$W_{10}$ and $P$, then it would be that the slope of the correlation
is steeper for faster rotating pulsars. Although the fit of a
second order polynomial through the data-points indeed show this
trend, it is statistically not much better than the first order fit.
High $\dot{E}$ pulsars are in general spinning faster then low
$\dot{E}$ pulsars, and therefore one could conclude that the pulse
widths of high $\dot{E}$ pulsars have a stronger dependence on $P$
than the low $\dot{E}$ pulsars. To illustrate this the pulsars with
high and low values of $\dot{E}$ are marked differently in
Fig. \ref{Fig_pulsewidths_p}. 
One could argue that there is not much evidence for a correlation for
the low $\dot{E}$ pulsars, while this is clearer for the high
$\dot{E}$ pulsars. But, as the fit second order polynomial was
statistically not much better than the fit of a power law, this
conclusion is also not significant. If this correlation exist, then
it would suggest that the profile widths of the high $\dot{E}$
pulsars follow the theoretical prediction more closely than those of
the low $\dot{E}$ pulsars, which could indicate that the emission
geometry for high $\dot{E}$ pulsars is more simple.

\subsection{Pulse widths versus $\dot{E}$}

\begin{figure}
\includegraphics[height=0.95\hsize,angle=270]{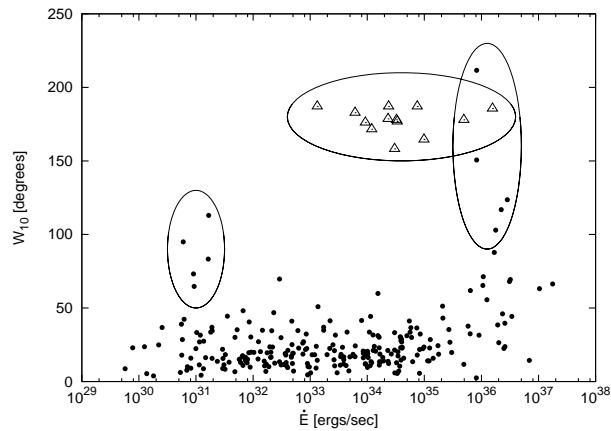}
\caption{\label{Fig_pulsewidths_edot}The measured profile 10\% widths
versus $\dot{E}$. The triangles indicate the pulsars which are
classified to have interpulses. Pulsars with a $S/N < 30$ and
profiles with substantial scattering were excluded. All the shown
observations were done at a wavelength of 20~cm. Three clusterings of
pulsars are highlighted by ellipses, which are discussed in the text.}
\end{figure}

In the previous subsection we found that $W_{10}$ is correlated with
$P$. One can expect that the correlation with $\dot{E}$ is weaker,
because $W_{10}$ was found to be uncorrelated with $\dot{P}$. Indeed,
Fig. \ref{Fig_pulsewidths_edot} shows that for most pulsars $W_{10}$
is as good as uncorrelated with $\dot{E}$. But remarkably, unlike in
Fig. \ref{Fig_pulsewidths_p} there are a number of outliers which are
clustered in relatively well defined regions in $\dot{E}$-space. These
outliers are indicated by the ellipses and each group will be
discussed separately below.

The first group of outliers are the pulsars in the ellipse at the left
hand side of Fig. \ref{Fig_pulsewidths_edot}. Although the profiles
are clearly wider than most pulse profiles, they form a continuous
distribution with the narrower profiles. These low $\dot{E}$ pulsars
are PSRs J1034--3224, J1655--3048 and J2006--0807 (which have complex
looking profiles) and PSRs J1133--6250 and J1137--6700 (which are
doubles with a clear saddle between the components).  The profiles of
these pulsars are most likely broad because their beam is close to
alignment with the rotation axis, making the beam intersect the line
of sight for a relatively long fraction of the rotation period. There
is evidence that the beam evolves to alignment with the rotation axis
over time (e.g. \citealt{wj08a}), so it is not surprising that these
aligned pulsars are old pulsars with low $\dot{E}$ values. For
two of these pulsars estimates for the angle between the magnetic axis
and the rotation axis can be found in the literature. These
polarization studies indeed suggest that the beam of PSRs J1034--3224
\citep{mhq98} and J2006--0807 \citep{ran93b,lm88} are close to
alignment.

The second group of pulsars with wide profiles are the pulsars with
interpulses, which are marked with triangles in
Fig. \ref{Fig_pulsewidths_edot}. These are PSRs J0834--4159,
J0905--5127, B0906--49, B1124--60, J1549--4848, B1607--52, B1634--45,
B1702--19, B1719--37, B1736--29, J1828--1101 and J1843--0702. The
profiles of these pulsars are characterised by having an interpulse
which is separated by approximately \degrees{180} in pulse longitude
from the main pulse. This separation is much larger than the widths of
the main- and interpulse. The most natural explanation for these
interpulses is that the emission of the main- and interpulse
originates from opposite magnetic poles. These pulsars are
concentrated to high values for $\dot{E}$. This is partially a
selection effect in the sample of pulsars which are included in the
Fermi timing program, but it has also shown by \cite{wj08a} that
interpulses are more likely to be detected in young (high $\dot{E}$)
pulsars.

The third group of pulsars with wide profiles can also be found at the
high $\dot{E}$ end of Fig. \ref{Fig_pulsewidths_edot}. These are PSRs
J1015--5719, B1259--63, J1803--2137, J1809--1917 and J1826--1334. Like
the group of pulsars with wide profiles at the low $\dot{E}$ end of
the figure, this group appears to form a continuum with the pulsars
with narrow profiles. We will refer to this group as the {\em
energetic wide beam pulsars}. Their profiles show a double structure
and they are exceptionally wide, but they are not separated by exactly
\degrees{180} in pulse longitude. In contrast to the group of
interpulses, this separation is not much larger than the width of the
individual components. The two components are often highly asymmetric
with steep edges at opposite sides, making the profile as a whole to
have a high degree of mirror symmetry. For some of these pulsars a
weak bump is detected in between the components, which disappears at
higher frequencies. The dependence of the PA on pulse longitude is
usually simple and straight.

It is not clear if PSR B1055--52 should be classified as an energetic
wide beam pulsar or a pulsar with an interpulse. On the one hand the
separation between the main- and interpulse is larger than the width
of the individual components, but on the other hand the components are
very wide and the interpulse is not exactly \degrees{180} away from
the main pulse. The location of PSR B1055--52 in
Fig. \ref{Fig_pulsewidths_edot} (the lowest triangle at
$\dot{E}=3.0\times10^{34}$ erg~s$^{-1}$) suggest that it is well
separated from the other energetic wide beam pulsars, although the
group of interpulse pulsars appear to have an overlap with the group
of energetic wide beam pulsars. Especially the location of PSRs
B0906--49 ($\dot{E}=4.9\times10^{35}$ erg~s$^{-1}$) and J1828--1101
($\dot{E}=1.6\times10^{36}$ erg~s$^{-1}$) in the figure are consistent
with both groups. Both these pulsars have interpulses at
$\sim$\degrees{180} away from the main pulse and this separation is
much larger than the component widths (the broad components of
J1828--1101 at 20~cm are because of scatter broadening), which is good
evidence that both interpulses are emitted from the opposite pole. For
PSR B0906--49 the PA-swing is shown to be inconsistent with a wide
cone interpretation (\citealt{kj08}).

The energetic wide beam pulsars are among the pulsars with the highest
$\dot{E}$ values ($\dot{E}>5\times10^{35}$ erg~s$^{-1}$), although it
is not true that all pulsars with high $\dot{E}$ values are also
energetic wide beam pulsars. This is first of all shown by the overlap
between the group of pulsars with interpulses and the energetic wide
beam pulsars group. Secondly, PSR J1513--5908 which has the highest
$\dot{E}$ value in our sample, does not show any evidence of a double
structure. Finally, PSR J1028--5819 is an extremely narrow double
(\citealt{kjk+08}, point in the bottom left corner of
Fig. \ref{Fig_pulsewidths_p}). A high $\dot{E}$ therefore appears to
be an important parameter which allows an energetic pulsar to form a
wide beam, but there must be more factors involved.

\subsection{The intensity ratio of the components of high $\dot{E}$ pulsars with double profiles}

\cite{jw06} noted that the trailing component of well separated
double profiles of high $\dot{E}$ pulsars tend to dominate in total
power (and in circular polarization as we will discuss below). This
curious effect seems to be strongest for pulsars with
$\dot{E}>10^{35}$ erg~s$^{-1}$ (see for example PSR J1420--6048,
Fig. \ref{J1420-6048}). The only exceptions are the Vela pulsar (which
has no well separated double profile), PSR J1302--6350 (an energetic
wide beam pulsars for which it is not clear which component is the
trailing component) and PSR J1831--0952. Nevertheless, in the majority
of the cases this correlation holds.

\section{Polarization}

\subsection{Linear polarization}

\begin{figure}
\includegraphics[height=0.95\hsize,angle=270]{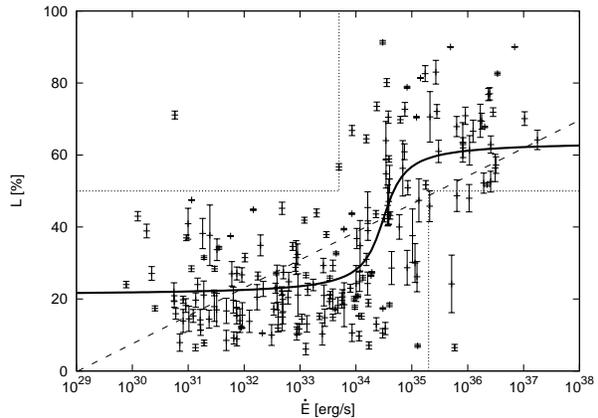}
\caption{\label{Fig_linearpol}The degree of linear polarization versus
$\dot{E}$ of all pulsars observed at 20~cm for which a significant
degree of linear polarization was measured. Pulsars which show
evidence for scatter broadening were excluded. There are two
relatively well defined regions which are almost empty in this
diagram. The dashed line shows the linear fit and the solid curve
the fit of an arctan function illustrating the step in the degree of
linear polarization.}
\end{figure}

\begin{figure}
\begin{center}
\includegraphics[width=0.95\hsize,angle=270]{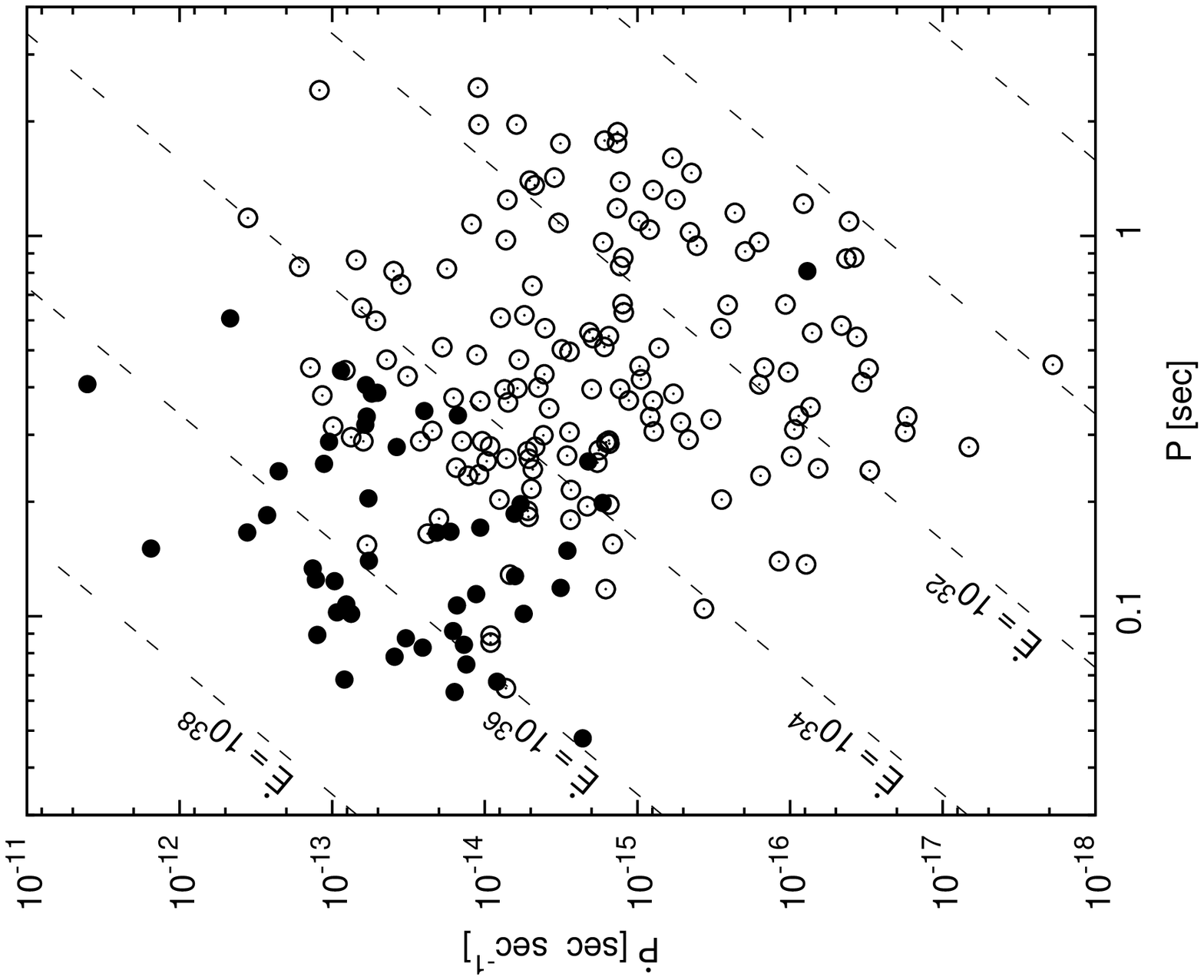}\\
\includegraphics[width=0.95\hsize,angle=270]{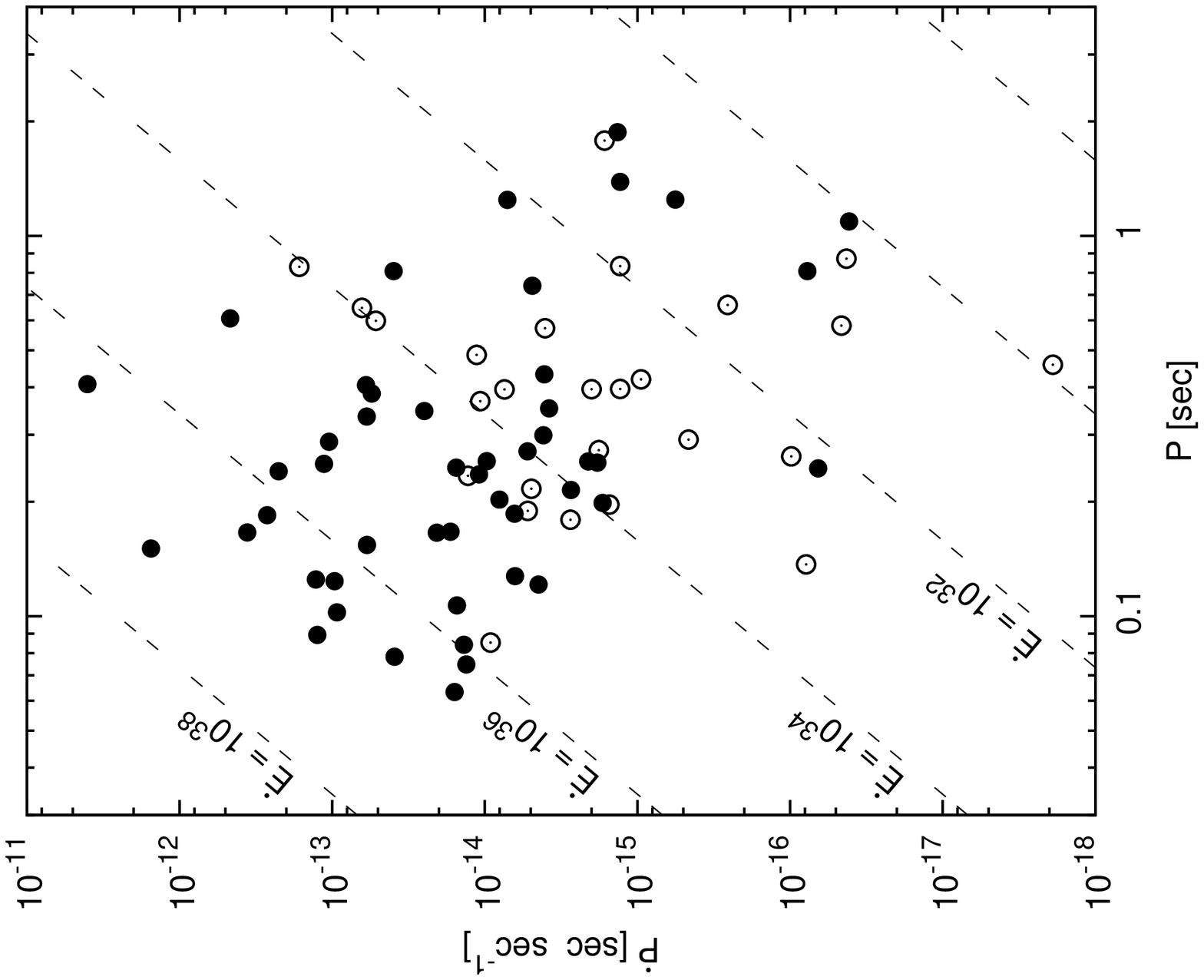}
\caption{\label{Fig_ppdot}{\em Top:} The $P-\dot{P}$ diagram of all
the observed pulsars at 20~cm for which a significant degree of linear
polarization could be measured. The filled and open circles are the
pulsars which are respectively more or less than 50\% linearly
polarized. {\em Bottom:} The $P-\dot{P}$ diagram of the pulsars which
show sudden jumps in their PA-swing (open circles) and those which
have a smooth PA-swing (filled circles). }
\end{center}
\end{figure}

It has been pointed out by several authors that degree of linear
polarization is high for high $\dot{E}$ pulsars
(e.g. \citealt{qmlg95,hlk98,cmk01,jw06}). This correlation is clearly
confirmed, as can be seen in Fig. \ref{Fig_linearpol}. There is a
transition from a low to a high degree of linear polarization which
happens around $\dot{E}\sim10^{34}-10^{35}$ erg~s$^{-1}$.  Virtually
all pulsars with $\dot{E} < 5\times10^{33}$ erg~s$^{-1}$ have less
than 50\% linear polarization and for almost all pulsars with with
$\dot{E} > 2\times10^{35}$ erg~s$^{-1}$ this percentage is above
50\%. There appears to be a transition region in between where
pulsars can both have low and high degrees of polarization, although
the transition is remarkably sharp and there are well defined spaces
in the figure which are almost empty. The non-linearity of the
degree of linear polarization versus $\dot{E}$ is confirmed by fitting
an arctan function through the data (solid curve in
Fig. \ref{Fig_linearpol}). This is done by minimizing the $\chi^2$
using the Levenberg-Marquardt algorithm \citep{mar63} as implemented
in \cite{ptvf92} (the data points are weighted equally). The total
$\chi^2$ is reduced by 20\% compared with a linear fit, which shows
that the step in the degree of linear polarization is important to
consider. Adding higher order polynomial terms does not reduce the
$\chi^2$ further, suggesting that the step is the most dominant
deviation from non-linearity. The position of the steepest point in
the fitted function occurs at $\log_{10}{\dot{E}} = 34.50\pm0.08$.

The emission of pulsars is thought to be a combination of two
orthogonally polarized modes (OPM, e.g. \citealt{mth75}). This aspect
of the emission can manifest itself in sharp $\sim\degrees{90}$ jumps
in the position angle (PA) over a small pulse longitude range. These
jumps are thought to be sudden transitions from the domination of one
mode to the other. Jumps in the PA-swing therefore indicates that both
modes are present in the emission. The mixing of both modes at a
certain longitude will lead to depolarization, so the presence of
jumps in the PA-swing could be anti-correlated with the degree of
polarization. By comparing the top and bottom panel of
Fig. \ref{Fig_ppdot} one can see that the $\dot{E}$ value at which the
transition from a low to a high degree of polarization takes place
coincides with the $\dot{E}$ value after which pulsars do not have
jumps in their PA. This is therefore important evidence that the
increase in the degree of linear polarization with $\dot{E}$ is caused
by one OPM dominating the emission. Most high $\dot{E}$ pulsars do not
show OPM jumps, but the reverse is not always true. Low $\dot{E}$
pulsars can have a low degree of linear polarization without evidence
for OPM jumps.

There are three curious exceptions in Fig. \ref{Fig_linearpol}
which do not follow the general trend. First of all PSRs
J1509--5850 and J1833--0827 have a low degree of linear polarization
while they have a high $\dot{E}$ ($5.2\times10^{35}$ and
$5.8\times10^{35}$ erg~s$^{-1}$ respectively). However it must be
noted that the leading and trailing components of PSR J1833--0827 are
highly polarized at 10~cm. The degree of linear polarization of
this pulsar shows a drop to zero in the middle of central component,
which could indicate that there is a transition in the dominating
OPM. The other exception is PSR J0108--1431, which has a low
$\dot{E}$ but is nevertheless highly polarized. This could suggest
that this pulsar has some similarities with high energy pulsars.

All pulsars without a significant amount of measured degree of linear
polarization fall below the $\dot{E} < 5\times10^{33}$ erg s$^{-1}$
line. The only exception is PSR J1055--6032, which appears to have a
very low degree of polarization. The rule that high $\dot{E}$ pulsars
are highly polarized therefore is confirmed in the majority of all
pulsars.

\subsection{Emission heights}

\subsubsection{The emission height derived from the pulse width}

The wider pulse profiles of high $\dot{E}$ pulsars are often
attributed to a larger emission height for those pulsars
(e.g. \citealt{man96,kj07}). The divergence of the magnetic (dipole)
field lines away from the magnetic axis makes the half opening angle
$\rho$ of the beam scale with the square root of the emission
height. Under the assumption that the beam of the pulsar is confined
by the last open field lines it follows that
\begin{equation}
\label{EqH}
\rho = \sqrt{\frac{9\pi \,\, h_{\rm em}}{2\,\,\, P\,\, c}}
\end{equation}
(e.g \citealt{lk05}), were $h_{\rm em}$ is the emission height and $c$
the speed of light.

\begin{table}
\begin{center}
\begin{tabular}{l c r r r r}
\hline
\multicolumn{1}{c}{Name} & $\dot{E}$ & $h_\mathrm{PA}$ & $h_{90}$ & \multicolumn{1}{c}{$\Delta\phi$} & $R_\mathrm{LC}$\\
 & [erg~s$^{-1}$] & [km] &  [km] & [deg] & [km]\\
\hline
J1015--5719 & $8.27\times10^{35}$ & --   & 5160 & -- & 6674\\
J1302--6350 & $8.25\times10^{35}$ & --   & 3476 & -- & 2279\\
J1803--2137 & $2.22\times10^{36}$ &   461 &  2967 &    8.3 &     6375 \\
J1809--1917 & $1.78\times10^{36}$ &   130 &  1425 &    3.8 &     3948 \\
J1826--1334 & $2.84\times10^{36}$ &   191 &  2522 &    4.5 &     4841 \\
\hline
J0304+1932 & $1.91\times10^{31}$ &    75 &   562 &    0.1 &    66206 \\
J0536--7543 & $1.15\times10^{31}$ &   -21 &  1472 &   -0.0 &    59444 \\
J0614+2229 & $6.24\times10^{34}$ &  1051 &    99 &    7.5 &    15982 \\
J0630--2834 & $1.46\times10^{32}$ &   118 &  2459 &    0.2 &    59376 \\
J0631+1036 & $1.73\times10^{35}$ &  1087 &   278 &    9.1 &    13731 \\
J0729--1448 & $2.81\times10^{35}$ &  1143 &   291 &   10.9 &    12008 \\
J0742--2822 & $1.43\times10^{35}$ &   338 &    68 &    4.9 &     7957 \\
J0835--4510 & $6.91\times10^{36}$ &    32 &    30 &    0.9 &     4263 \\
J0908--4913 & $4.92\times10^{35}$ &   320 &    24 &    7.2 &     5094 \\
J1048--5832 & $2.01\times10^{36}$ &   -24 &   141 &   -0.5 &     5901 \\
J1105--6107 & $2.48\times10^{36}$ &   243 &    52 &    9.2 &     3015 \\
J1119--6127 & $2.34\times10^{36}$ &  2082 &  1406 &   12.3 &    19455 \\
J1123--4844 & $1.76\times10^{32}$ &   540 &   152 &    5.3 &    11682 \\
J1253--5820 & $4.97\times10^{33}$ &   690 &   116 &    6.5 &    12191 \\
J1320--5359 & $1.67\times10^{34}$ &   673 &   158 &    5.8 &    13347 \\
J1359--6038 & $1.21\times10^{35}$ &   544 &    41 &   10.2 &     6084 \\
J1420--6048 & $1.04\times10^{37}$ &   106 &   442 &    3.7 &     3253 \\
J1531--5610 & $9.09\times10^{35}$ &    81 &   136 &    2.3 &     4018 \\
J1535--4114 & $1.98\times10^{33}$ &   474 &   235 &    2.6 &    20654 \\
J1637--4553 & $7.51\times10^{34}$ &   392 &    62 &    7.9 &     5667 \\
J1701--3726 & $2.97\times10^{31}$ &   184 &   984 &    0.2 &   117118 \\
J1705--3950 & $7.37\times10^{34}$ &  -120 &   690 &   -0.9 &    15218 \\
J1709--4429 & $3.41\times10^{36}$ &   563 &   326 &   13.2 &     4889 \\
J1733--3716 & $1.54\times10^{34}$ &   932 &  1969 &    6.6 &    16107 \\
J1740--3015 & $8.24\times10^{34}$ &   818 &    31 &    3.2 &    28952 \\
J1835--1106 & $1.78\times10^{35}$ &   445 &    93 &    6.4 &     7916 \\
J1841--0345 & $2.69\times10^{35}$ &   353 &   415 &    4.2 &     9737 \\
\hline
\end{tabular}
\end{center}
\caption{\label{emissionheights90}The emission height $h_\mathrm{PA}$
is derived from the offset $\Delta\phi$ between the inflection point
of the PA-swing and the centre of the pulse profile. The emission
height $h_{90}$ is derived from the pulse width assuming an orthogonal
rotator ($\alpha=\degrees{90}$) and a line of sight which makes a
central cut through the emission beam ($\beta=\degrees{0}$), which
is the emission height for a typical random geometry. The last column
is the light cylinder radius. The first five pulsars are the pulsars
which we have classified as energetic wide beam pulsars.}
\end{table}

Wider beams are more likely to produce wide profiles, although
the observed pulse width also depend on the orientation of the
magnetic axis and the line of sight with respect to the rotation
axis. The relevant parameters are the angle $\alpha$ between the
magnetic axis and the rotation axis and the angle $\zeta$ between the
line of sight and the rotation axis. A related angle is the impact
parameter $\beta = \zeta-\alpha$, which is the angle between the line
of sight and the magnetic axis at its closest approach. For most
pulsars it is extremely difficult to obtain reliable values for these
angles, which makes it hard to derive the emission height from
$W_{10}$.

For a sample of pulsars with a random orientation of the magnetic axis
and the line of sight both the $\alpha$ and $\zeta$ distribution are
sinusoidal. Simulations using the model described in \cite{wj08a},
show that the pulse width distribution for such a sample pulsars peaks
at $2\rho$. Some pulsars will have wider profiles because the pulsar
beam is more aligned with the rotation axis, while others will have
narrower profiles because the line of sight grazes the beam. This
implies that the typical pulse width of a large sample of pulsars
which have random orientations of their spin and magnetic axis and
have similar opening angles $\rho$ should be equal to $2\rho$. In
other words, a typical profile width is equal to that which is
expected for an orthogonal rotator ($\alpha=\degrees{90}$) and a line
of sight which makes a central cut through the emission beam
($\beta=\degrees{0}$). For such geometry Eq. \ref{EqH} can be
rewritten to
\begin{equation}
\label{EqH90}
h_{90} = \frac{cP\left(W_{10}\right)^2}{18\pi},
\end{equation}
which is the emission height for a typical random geometry assuming a
magnetic dipole field and an active area of the polar cap which is set
by the last open field lines.

\subsubsection{The emission height derived from the PA-swing}

An independent way to estimate the emission height is by measuring the
shift of the PA-swing caused by the co-rotation of the emission region
with the neutron star. In this method it is assumed that the PA-swing
is described by the rotating vector model (RVM; \citealt{rc69a}). The
position angle $\psi$ is then predicted to depend on the pulse
longitude $\phi$ as
\begin{equation}
\tan\left(\psi-\psi_0\right)=\frac{\sin\alpha\;\sin\left(\phi-\phi_0\right)}{\sin\zeta\;\cos\alpha-\cos\zeta\;\sin\alpha\;\cos\left(\phi-\phi_0\right)},
\end{equation}
where $\psi_0$ and $\phi_0$ are the PA and the pulse longitude
corresponding to the intersection of the line of sight with the
fiducial plane (the plane containing the rotation and magnetic axis).
The PA-swing is a S-shaped curve and its inflection point occurs at
$\phi_0$. The RVM fit is shown in the figures of Appendix A and B for
the pulsars which have a roughly S-shaped PA-swing.

If the emission profile is symmetric around the magnetic axis, then one
could expect the inflection point to coincide with the middle of the
pulse profile. However, co-rotation causes the inflection point to be
delayed with respect to the pulse profile. The pulse longitude
difference $\Delta\phi$ between the middle of the profile and the
inflection point of the PA-swing can be used to derive the emission
height (\citealt{bcw91})
\begin{equation}
\label{heighBCW92}
h_\mathrm{PA} = \frac{P\,c\,\Delta\phi}{8\pi }.
\end{equation}
The relative shift of the PA-swing with respect to the profile is
independent of $\alpha$ and $\zeta$ \citep{drh04}. If the
emission height is too large it could be difficult to measure
$\Delta\phi$ because the inflection point of the PA-swing is shifted
beyond the edge of the pulse profile.

\subsubsection{The derived emission heights}

\begin{figure}
\includegraphics[height=0.95\hsize,angle=270]{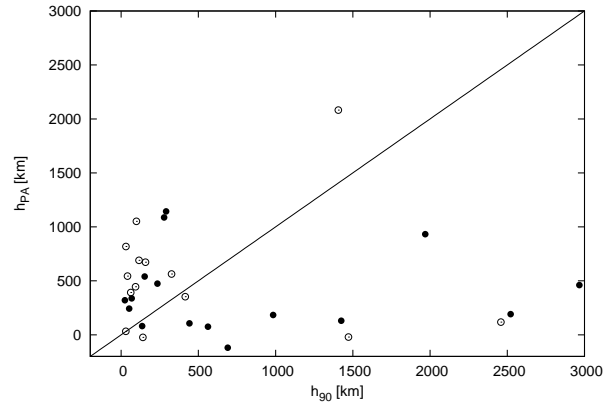}
\caption{\label{Fig_height}The emission height as derived using two
independent methods, one using the pulse widths ($h_{90}$) and the
other using the shift of the PA-swing with respect to the profile
($h_\mathrm{PA}$). The solid circles indicate profiles with a clear
double structure. The points should lie on the line if the emission
heights are consistent with each other.}
\end{figure}

The emission height derived using the PA-swing ($h_\mathrm{PA}$) and
using the pulse width ($h_\mathrm{90}$) are both listed in Table
\ref{emissionheights90}. This only includes the pulsars which have a
clear S-shaped PA-swing at 20~cm. The typical emission height is a few
hundred km, which is similar to the emission height found by other
authors (e.g. \citealt{bcw91,mr01}).  One can see that for some
pulsars $h_\mathrm{PA}$ is negative, which is obviously
impossible. This can be considered to be a clear warning that the
emission heights for an individual source could be completely wrong,
but one can nevertheless hope that they are meaningful in a
statistical sense. In order to test this we calculated the Spearman
rank-order correlation coefficient between $h_\mathrm{PA}$ and
$h_\mathrm{90}$, which shows that there is no evidence for any
correlation between these parameters. This is also evident from
Fig. \ref{Fig_height}, where these quantities are plotted against each
other. We are therefore forced to accept that even in a statistical
sense the calculated emission heights are inconsistent, supporting
the same conclusion reached by \cite{ml04} based on six pulsars.

There are a number of reasons why the heights derived using the
two methods could be inconsistent. If the beams are significant
patchy, then the centroid of the profile is not related to the
position of magnetic axis and both methods to derive emission heights
will fail. We therefore made a distinction in Fig. \ref{Fig_height}
between the profiles which are clear doubles and other profiles,
because the double structure could indicate that the pulsar beam is
roughly symmetric around the magnetic axis. As one can see there is no
noticeable difference in the distributions. Another effect that could
be important is the effect of sweepback of the magnetic field
lines. \cite{dh04} derived that the effect of sweepback can dominate
over other effects of co-rotation at low altitudes, making it possible
for the inflection point of the PA-curve to precede the profile
centre. The PA-curve can also precede the profile in case of inward
directed emission (\citealt{dfs+05}).

Despite the inconsistency between the derived emission heights using
both methods, it is not true that the emission heights are entirely
random. Most pulsars show a positive emission height $h_\mathrm{PA}$,
indicating that the steepest slope of the PA-swing trails the centroid
of the profile in most cases. In fact, Fig. \ref{Fig_height}
appears to show evidence that it is unlikely that both $h_{90}$ and
$h_\mathrm{PA}$ are large. In Table \ref{emissionheights90} one can
see that the emission of the energetic wide beam pulsars should come
from near the light cylinder in order to explain the width of the
pulse profiles ($h_\mathrm{90}\sim R_\mathrm{LC}$). However, the
derived emission heights from the the PA-swing fits are not unusually
large. In this list one could add the emission height of PSR
J1015--5719, which is estimated by \cite{jw06} to be 380 km. 

All the energetic wide beam pulsars with a derived emission height
from the PA-swing can be found below the solid line in
Fig. \ref{Fig_height}, as well as PSRs J1705--3950 and J1733--3716
which have similar profile shapes. It seems unlikely that they all
have beams which are close to alignment with the rotation axis,
which suggest a different reason for the large widths of the profiles
of the energetic wide beam pulsars.  Apparently the emission heights
which are derived from the PA-swings of the energetic wide beam
pulsars are systematically underestimated, or the heights derived from
the profile widths are over estimated. The first case could be
explained by magnetic field line sweepback when the emission height is
low (\citealt{dh04}). The second case implies that the beams of these
pulsars are wider than could be expected from the divergence of the
dipole field lines. The widening of the pulsar beam could, at least in
principle, be caused by propagation effects in the magnetosphere.

Another explanation for the deviation of the energetic wide beam
pulsars from the line in Fig. \ref{Fig_height} could be that the two
methods estimate the emission heights at different locations in the
magnetosphere. The method based on the profile width estimates the
emission height at the edge of the of the beam, while the method using
the PA-swing estimates the emission height of the more central
regions of the beam. For the energetic wide beam pulsars $h_{90}$ was
found to be systematically larger than $h_\mathrm{PA}$, which can be
interpreted as evidence for an increase in the emission height at the
edge of the beam. This interpretation will be discussed in more detail
in the following section.

Fig. \ref{Fig_height} shows that besides the group of pulsars which
have relatively large $h_{90}$ compared to $h_\mathrm{PA}$, there is
also a group in where the opposite is seen. An explanation could be
that for those pulsars only a fraction of the polar cap is active 
(e.g. \citealt{kg97}). Support for this interpretation is that the
profiles of a number of pulsars in this group are argued to be
produced by partial cones, including PSR J0543+2329
(\citealt{wck+04}), J0614+2229 (\citealt{jkk+07}) and J0659+1414
(\citealt{ew01}).

\subsection{Circular polarization}

\begin{figure}
\includegraphics[height=0.95\hsize,angle=270]{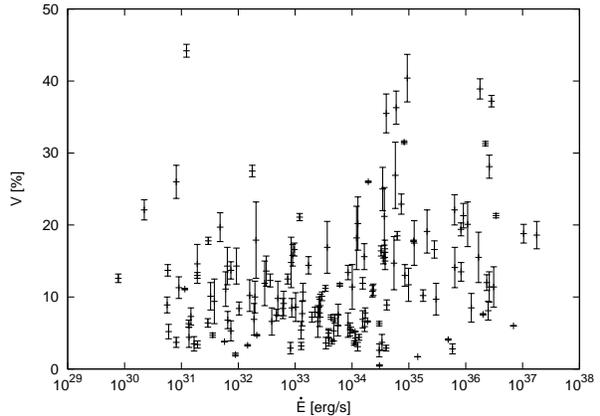}
\caption{\label{Fig_circularpol}The degree of circular polarization
(the absolute value of Stokes V) versus $\dot{E}$ of all pulsars
observed at 20~cm for which a significant degree of circular
polarization was measured. Pulsars which show evidence for scatter
broadening were excluded. }
\end{figure}

Unlike the degree of linear polarization the degree of circular
polarization appears to be unaffected by $\dot{E}$. Also the fraction
of pulsar which are left- and right-hand circularly polarized is about
50 per-cent for both the high and low $\dot{E}$ pulsars.

\cite{jw06} noted that, besides the total intensity, also the
degree of circular polarization usually dominates in the trailing
components of high $\dot{E}$ pulsars with well separated double
profiles. This correlation is also clearly confirmed in our data for
all pulsars with an $\dot{E}>10^{34}$ erg s$^{-1}$. The only clear
counter example in our data-set could be PSR J1705--1906
($\dot{E}=7\times10^{34}$ erg s$^{-1}$), which has a high degree of
circular polarization in the leading half of the profile. However, it
must be noted that the single pulse modulation properties of this
pulsar suggest that the leading component is not the leading component
of a double, but rather a precursor to a blended double which forms
the trailing half of the profile (\citealt{wws07}).

Remarkable correlations have been reported between the sign of the
circular polarization and the sign of the slope of the
PA-swing. According to \cite{rr90} the sign of the slope of the
PA-swing is correlated with the sign of the circular polarization for
pulsars which are cone dominated and for which the sign of the
circular polarization is the opposite in the two components. But
\cite{hmxq98} did not confirm this correlation. Instead they propose
that there is a correlation between the sign of the circular
polarization and the sign of the slope of the PA-swing for cone
dominated pulsars which have the same sign of circular
polarization. Our data does not show much evidence for either
correlation. Compare for instance the plots for PSR J1826--1334
(positive circular polarization, decreasing PA-swing) with PSR
J1733--3716 (negative circular polarization, decreasing PA-swing).

\section{Discussion}

\subsection{Extended radio emission regions?}

We concluded that the differences in the pulse profile morphology of
the high and low $\dot{E}$ pulsars is in general rather subtle,
without an objectively measurable discriminator between them. An
exception is what we call the group of energetic wide beam pulsars,
which do have distinct profile properties and which will be discussed
separately below.  The measured slope of the $W_{10}-P$ correlation
appears to be flatter than the theoretical $P^{-1/2}$ slope.  It is
far from straight-forward to link the deviation of the slope to a
physical mechanism. For example, as explained in \cite{wj08a}, the
measured slope depends on the details of the evolution of the pulsar
spin-down and the alignment of the magnetic axis with the rotation
axis. If the active area of the polar cap is influenced by other
factors than just the opening angle of the last open field lines, or
if the emission height varies from pulsar to pulsar, one can expect to
observe its effects in the $W_{10}-P$ correlation. There is some
marginal evidence that the slope of the correlation is steeper for
high $\dot{E}$ pulsars, suggesting that for high $\dot{E}$ pulsars
these other factors are less important.

If one believes that the emission geometry is simpler for high
$\dot{E}$ pulsars, one can ask the question of what is causing
this. One factor that could affect the complexity of profiles is the
emission height. Complexity could arise because of multiple distinct
emission heights within the beam (\citealt{kj07}). In their model high
$\dot{E}$ pulsars have only one emission height which is similar for
different pulsars. However, one could also make the argument for an
opposite effect. Maybe the emission of high $\dot{E}$ pulsars is not
emitted from a well defined height, but rather from an extended height
range. This would mean that the observed profiles of high $\dot{E}$
pulsars are a superposition of profiles emitted from a continuum of
heights. The observed sum of those profiles (shifted with respect to
each other by aberration and retardation) will have less complexity
because they are blurred out. Not only are the profiles expected to be
less complex in this scenario, but there is also not much room to vary
the emission height from pulsar to pulsar if the height range is
large. This would make the $W_{10}-P$ correlation follow the
prediction more closely. Large emission height ranges are typical for
high energy models, such as the slot gap models (e.g. \citealt{mh04})
or the two pole caustic models (\citealt{dr03}), hence there could be
parallels with the radio emission for high $\dot{E}$ pulsars. These
parallels could be even more relevant for the energetic wide beam
pulsars.

\subsection{The energetic wide beam pulsars}

As discussed by e.g. \cite{man96}, there is a group of young pulsars
which can be found among the highest $\dot{E}$ pulsars which have very
wide profiles with often steep edges. The profiles are clearly mirror
symmetric, suggesting that the components are the two sides of a
single beam rather than two beams from opposing magnetic poles. This
interpretation is also suggested by the frequency evolution of PSR
B1259--63 (\citealt{mj95}). Because these objects are young, the
typical orientation of the magnetic axis is not expected to be very
different for the highest $\dot{E}$ pulsars and those with
intermediate values, which suggests that some pulsars with high
$\dot{E}$ values can have very different beams compared with other
pulsars.

An interesting analogue can be drawn between the radio profiles of
energetic wide beam pulsars and the high energy profiles of
pulsars. High energy profiles can also be wide doubles which often
have sharp edges (e.g. \citealt{tho04}).  The pulsars which produce
high energy emission are the pulsars with high $\dot{E}$ values, so
there could be a direct link between the high energy pulsars and the
energetic wide beam pulsars. Maybe the radio emission and the high
energy emission are produced at the same location in the
magnetosphere. The sharp edges of the high energy profiles are often
explained by caustics which form because of the combined effect of
field line curvature, aberration and retardation
(e.g. \citealt{mor83}). These caustics occur when the emission is
produced high in the magnetosphere over a large altitude range and if
the magnetic axis is not aligned to the rotation axis
(e.g. \citealt{dr03}). If the radio emission and $\gamma$-ray emission
would come from similar locations, one would expect the radio and
$\gamma$-ray profiles to look alike. Hopefully the Fermi satellite
will find high energy counterparts for these pulsars which allows a
test of this hypothesis.

Another way to produce profiles with sharp edges could be the
combination of refraction of radio waves in pulsar magnetospheres in
combination with an emission height which is different for different
field lines (\citealt{wsv+03}). Only the ordinary wave mode is
refracted (e.g. \citealt{ba86}) or scattered (\citealt{pet08a}) in the
magnetosphere. The profiles of the energetic wide beam pulsars can
therefore be expected to be dominated by one polarization mode, which
could potentially also explain their high degree of linear
polarization. The unpolarized bump which is observed in the middle of
some of these profiles could be the un-refracted part of the beam,
which is depolarized because the presence of the extra-ordinary mode.
These central components are strongest at lower frequencies,
consistent with the steeper spectral index which is often observed for
the central components of pulse profiles (e.g. \citealt{ran83}). 
However, it remains to be seen if propagation effects can be strong
enough to explain the extreme pulse widths which are observed.

The emission geometry appears to be different for high $\dot{E}$
pulsars and is possibly more similar to that of the high energy
emission. However, not all pulsars with high $\dot{E}$ values produce
these extremely wide profiles. Apparently only a subset of the high
$\dot{E}$ pulsars have emission geometries which are very different
from normal radio pulsars. A high $\dot{E}$ is therefore an important
parameter required for the energetic wide beam pulsars, but not the
only one. For instance, maybe only certain configurations of the
plasma distributions enlarge the beam via propagation effects or maybe
not all pulsars have a slot gap which produces radio emission. 
It must also be noted that, like the high $\dot{E}$ radio pulsars, not
all the high energy pulse profiles of pulsars are doubles (e.g. PSR
B1706--44).

\subsection{Emission heights}

There is no evidence that pulsars with large emission heights (derived
from their PA-swing) have wider profiles. It is therefore not clear
what the physical meaning of these emission heights is. There are many
reasons why the derived emission heights could be wrong, including
asymmetric beams, partially active polar caps or sweepback of the
magnetic field lines.  Also if the PA-swing of emission which is
emitted far out in the magnetosphere the PA can be expected to deviate
from the rotating vector model. For instance, the PA-swing for the
outer gap model is predicted to have the steepest slope near the edges
of the profile (\citealt{ry95}), rather then at the pulse longitude
corresponding to the location of the magnetic axis. This would
complicate the calculation of emission heights from the observed
PA-swing considerably. Nevertheless, the fact that most PA-swings
trail the centroid of the profile suggest that the derived emission
heights do carry some information.

The emission of the energetic wide beam pulsars should come
from near the light cylinder in order to explain the width of the
pulse profiles. However, the derived emission heights from the
PA-swing fits seem to suggest that the emission heights are not
unusually large. This can be seen as support that the beams of
energetic wide beam pulsars are wide because of propagation effects
instead of caused by a large emission height. An alternative
interpretation is that the emission height at the edge of the beam is
much larger than in the centre of the beam. This would fit in nicely
with the result of \cite{gg03}, who concluded that the outer
components of PSR B0329+54 are emitted from higher in the
magnetosphere. It also fits in nicely with the hypothesis that the
emission of the energetic wide beam pulsars comes from an extended
emission height range, making the emission geometry very similar to
the slot gap model.

\subsection{Interpulse problem?}

The conclusion that the beams of energetic wide beam pulsars are large
appears to be unavoidable. If this is the case, than one would expect
that it is very likely for the line of sight to intersect the beams of
both poles of the pulsar. Using the model described by \cite{wj08a}
the probability for the line of sight to intersect both beams is
predicted to be 64\%, assuming $\rho=\degrees{75}$ and a random
orientation of the magnetic axis and the line of sight. However, there
is no clear example of an energetic wide beam pulsar which has a
(double peaked) interpulse.
The ``interpulse problem'' is then why we do not observe the
interpulses of the energetic wide beam pulsars.

It is argued by \cite{man96} that the steep edges form the outer edge
of an extremely wide beam, which would make the peak-to-peak
separation of the profiles wider than \degrees{180}. In that case the
weak bumps observed for some of these pulsars are then separated by
half a rotational phase from the centre of the profile, which would
make them interpulses. However, because the bumps fill in the region
in between the sharp edges, it seems more likely that the sharp edges
form the inner edges of a wide beam. In that case the profiles are
less wide and the bump forms the centre of the same wide beam.

The interpulse problem suggests that the beam sizes are different
for the magnetic poles of the energetic wide beam pulsars. As
discussed above, not all profiles of high $\dot{E}$ pulsars have wide
components. This implies that other criteria have to be met in order
to make the beams wide. These criteria are not necessarily met
simultaneously for both poles, which would reduce the fraction of
pulsars with interpulses. The very different shape of the main- and
interpulse of PSR B1055--52 show that interpulse beams can have very
different shapes, hence possibly also very different sizes. Only 5 of
the 26 pulsars with an $\dot{E}>5\times10^{35}$ erg~s$^{-1}$ in
Fig. \ref{Fig_pulsewidths_edot} are classified to be energetic wide
beam pulsars. Therefore the chance that both poles produces a wide
beam is expected to be only $\sim4\%$ if the chance of producing a
wide beam is independent for each pole.

A more extreme point of view to solve the interpulse problem is put
forward by \cite{man96} who argues, following \cite{ml77}, that all
pulsars only have one active wide beam. Although this would trivially
solve the interpulse problem, it does not explain the concentration of
main- interpulse separations near \degrees{180} (see
Fig. \ref{Fig_pulsewidths_edot}).

\subsection{Polarization}

The degree of linear polarization is found in several studies to
increase with $\dot{E}$. Such behaviour is predicted for the natural
wave modes in the cold plasma approximation (\citealt{hlk98}).  One of
the most surprising results of this paper is the sudden increase in
the degree of linear polarization with $\dot{E}$. This suggest that
pulsars can be separated into two groups which have distinct physical
properties. This could either be in the structure of the magnetosphere
or the physics of the emission mechanism itself. It is remarkable that
over 7 orders of magnitude in $\dot{E}$ the degree in linear
polarization is the only thing that is clearly changing.

It has been shown that the degree of polarization is clearly related
to the presence of OPM transitions in the PA-swing. The two plasma
modes (X-mode and O-mode) can be expected to be separated more in
pulse longitude for high $\dot{E}$ pulsars, because the difference in
their refractive indices is larger (e.g. \citealt{hlk98}). This could
prevent the modes from mixing, and therefore prevent
depolarization. However, the fact that high $\dot{E}$ pulsars are less
likely to show jumps in their PA-swing suggests that they only
effectively generate one of the modes. \cite{jhv+05} found that the
velocity vectors of most pulsars make an angle close to either
\degrees{0} or \degrees{90} with the PA of the linear polarization
(measured at the inflection point). This is interpreted as evidence
for alignment of the rotation axis of the star with its proper motion
vector and the bimodal nature of the distribution of angles is
interpreted to be due to the domination of different plasma modes for
different stars. This result therefore suggest that if the emission of
high $\dot{E}$ pulsars is dominated by one mode, it could be either of
the two for different pulsars.
If the profiles of the energetic wide beam pulsars are widened by
refraction, than their emission should be dominated by the O-mode
which can be refracted in the pulsar magnetosphere.

A very different interpretation of the sudden increase of the degree
of linear polarization with $\dot{E}$ is based on the fact that the
$\dot{E}$ at the transition is very similar to the death line for
curvature radiation (\citealt{hm02}). This death line could
potentially cause a sudden change in for instance the plasma
distribution in the magnetosphere (which is responsible for the
refraction of the plasma waves), or it could possibly change the
emission mechanism which is responsible for the radio emission. The
possible link between the degree of linear polarization of the radio
emission and the mechanism for the production of the high energy
emission could therefore suggest that the $\gamma$-ray efficiency is
correlated with the degree of linear polarization in the radio band.
It would therefore be extremely interesting to find out if a pulsar
like PSR J0108--1431, which is highly polarized with a low $\dot{E}$,
can be detected by the Fermi satellite.

\subsection{Circular polarization}

The trend noted by \cite{jw06} that the degree of circular
polarization is usually higher in the second component of doubles is
clearly present in our data as well. It is a possibility that this is
a result of the co-rotation velocity of the emission region. As shown
by for instance \cite{dyk08}, particles travelling along the magnetic
field lines of a rotating dipole will follow stronger curved paths (in
the inertial observer frame) at the leading half of the pulse profiles
compared to the trailing half of the pulse profile. This is the reason
why the observed PA-swing appears to be shifted with respect to the
pulse profile (equation \ref{heighBCW92}). The degree of circular
polarization is in general highest in the central parts of the pulse
profile, there where the curvature of the field lines is weakest. This
could therefore suggest that the location of the highest degree of
circular polarization is, like the PA-swing, shifted to later times by
co-rotation.

\section{Conclusions}

In this paper we present and discuss the polarization profiles of a
large sample of young, highly energetic pulsars which are regularly
observed with the Parkes telescope. This sample is compared with a
sample of a similar number of low $\dot{E}$ objects in order to draw
general conclusions about their differences.

There is some evidence that the total intensity profiles of high
$\dot{E}$ pulsars are slightly simpler based on a classification by
eye. However, there is no difference in the complexity of the
mathematical decomposition of the profiles, the amount of overlap
between the components of doubles or the degree of profile symmetry.
We therefore conclude that differences in the total intensity pulse
morphology between high and low $\dot{E}$ pulsars are in general
rather subtle. High $\dot{E}$ pulsars appear to show a stronger
$W_{10}-P$ correlation which is closer to the theoretical expectation,
suggesting that for high $\dot{E}$ pulsars there are less complicating
factors in the emission geometry.

A much more pronounced difference between high and low $\dot{E}$
pulsars is the degree of polarization. The degree of polarization was
already known to increase with $\dot{E}$, but our data shows there is
a rapid transition between relatively unpolarized low $\dot{E}$
pulsars and highly polarized high $\dot{E}$ pulsars. The increase in
the degree of polarization is related to the absence of OPM
jumps. Refraction of the radio emission is expected to be more
effective in the magnetosphere of high $\dot{E}$ pulsars, which could
prevent depolarization because of mixing of the plasma modes. The
absence of OPM jumps suggest that one of the two modes (not
necessarily the same for different pulsars) dominates over the other.
The $\dot{E}$ of the transition is very similar to the death line for
curvature radiation, which could be the reason why the transition is
relatively sharp. This potential link between the high energy
radiation and the radio emission could mean that the $\gamma$-ray
efficiency is correlated with the degree of linear polarization in the
radio band.

The degree of circular polarization is in general higher in the second
component of doubles. This remarkable correlation is possibly caused
by the effect of co-rotation on the curvature of the field lines in
the inertial observer frame, making this effect very similar to the
shift of the PA-swing predicted for a finite emission height. In
addition, the trailing component usually dominates in total power.

The $W_{10}-\dot{E}$ distribution clearly shows sub-groups which are
not visible in the pulse $W_{10}-P$ distribution, suggesting that
$\dot{E}$ is an important physical parameter for pulsar
magnetospheres. Besides a group of pulsars which probably have beams
aligned with the rotation axis and a group of pulsars with interpulses
which are probably orthogonal rotators there is a group of energetic
wide beam pulsars. These young pulsars have very wide profiles with
often steep edges which are likely to be emitted from a single pole.

The profile properties of the energetic wide beam pulsars are similar
to those of the high energy profiles, suggesting another possible link
with the high energy emission. We therefore propose that the emission
of these pulsars could come, like the high energies, from extended
parts of the magnetosphere. 
The extended height range from where the emission is emitted will
smear out the complex features of the profiles.
A large height range could also prevent the emission height to vary
much from pulsar to pulsar, which would result in a stronger
$W_{10}-P$ correlation for high $\dot{E}$ pulsars, as is indeed
observed. If the radio emission and $\gamma$-ray emission of these
pulsars indeed come from similar locations in the magnetosphere, one
would expect the radio and $\gamma$-ray profiles to look alike,
something that potentially can be tested by the Fermi satellite.

An alternative mechanism to produce the profiles of the energetic wide
beam pulsars could be the combination of refraction (or scattering) of
radio waves in pulsar magnetospheres with an emission height which is
different for different field lines (\citealt{wsv+03}). Refraction
(and scattering) is most severe for the ordinary wave mode, suggesting
that these profiles are dominated by one polarization mode. This would
be consistent with the high degree of linear polarization observed for
these pulsars. The unpolarized bump in the middle of the profiles of
the energetic wide beam pulsars could be the un-refracted part of the
beam, which is depolarized because the mixing of the plasma modes.

Measurements of the emission height could potentially discriminate
between the refraction model and the extended emission height model
for the energetic wide beam pulsars. There is no evidence that pulsars
with large emission heights (derived from their PA-swing) have wider
profiles. It is therefore not clear what the physical meaning of these
emission heights are. It could supports the idea that the beams of
energetic wide beam pulsars are wide because of refraction instead of
caused by a large emission heights. However, it could also mean that
the emission height of the outer parts of the beam is much larger than
for the central parts, making the emission geometry similar to that of
a slot gap.

\section*{Acknowledgments}
The authors would like to thank the referee, Axel Jessner, for his
useful comments on this paper. The Australia Telescope is funded by
the Commonwealth of Australia for operation as a National Facility
managed by the CSIRO.

\bibliographystyle{mn2e}
\bibliography{journals_apj,modrefs,psrrefs} 
\label{lastpage} 

\appendix
\section{Polarization figures at both 10 and 20~cm}
{\bf The astro-ph version is missing 528 figures due to file size
restrictions. Please download the paper including the appendices from
http://www.atnf.csiro.au/people/pulsar/wj08b.pdf.}
\section{Polarization figures only at 20~cm}
{\bf The astro-ph version is missing 528 figures due to file size
restrictions. Please download the paper including the appendices from
http://www.atnf.csiro.au/people/pulsar/wj08b.pdf.}
\section{Table with the profile properties}
\onecolumn
\input{table.tex}

\clearpage
\end{document}

%% file: table.tex
\setcounter{table}{0}
\begin{table*}
\begin{tabular}{lcrcrcrrcccrc}
\multicolumn{1}{c}{Name} & $\lambda_\mathrm{obs}$ & \multicolumn{1}{c}{$t_\mathrm{obs}$} &  Class    &  \multicolumn{1}{c}{$S/N$} & Scat. &  \multicolumn{1}{c}{$W_{50}$}  &  \multicolumn{1}{c}{$W_{10}$}    &  $N_\mathrm{Comp}$    &  Sym.  &  \multicolumn{1}{c}{$L$}  &  \multicolumn{1}{c}{$V$}  & Fig. \\
 & [cm] & \multicolumn{1}{c}{[sec]} & & & & \multicolumn{1}{c}{[\degr]} & \multicolumn{1}{c}{[\degr]} & & & \multicolumn{1}{c}{[\%]} & \multicolumn{1}{c}{[\%]}\\
\hline
\hline
J0034--0721     &        20   & 959 &  S   &     217  &           N  & 12.4 & 34.5 &        2  &   0.964  &   7.8 $\pm$   0.8  &    3.4 $\pm$   0.5  & \ref{J0034-0721} \\
J0051+0423     &        20   & 3838 &  D   &      32  &           N  & 31.1 & 42.3 &  &   0.926  &                 &                  & \ref{J0051+0423} \\
J0108--1431     &        10   & 6565 &  S   &      54  &           N  & 10.5 & 26.3 &  &   0.978  &  40.0 $\pm$   3.9  &                  & \ref{J0108-1431} \\
J0108--1431     &        20   & 19731 &  S   &     186  &           N  & 12.2 & 28.5 &        2  &   0.990  &  71.1 $\pm$   1.1  &   13.7 $\pm$   0.8  & \ref{J0108-1431} \\
J0134--2937     &        20   & 960 &  D   &     238  &           N  & 5.6 & 18.2 &        4  &   0.803  &  41.9 $\pm$   0.8  &  -21.1 $\pm$   0.5  & \ref{J0134-2937} \\
J0151--0635     &        20   & 959 &  D   &      51  &           N  & 29.3 & 39.0 &  &   0.884  &  22.0 $\pm$   2.5  &                  & \ref{J0151-0635} \\
J0152--1637     &        20   & 958 &  D   &     506  &           N  & 6.8 & 9.8 &        3  &   0.939  &  12.1 $\pm$   0.3  &   -2.0 $\pm$   0.2  & \ref{J0152-1637} \\
J0206--4028     &        20   & 960 &  D   &      52  &           N  & 3.5 & 10.1 &  &   0.973  &                 &    6.9 $\pm$   2.2  & \ref{J0206-4028} \\
J0211--8159     &        20   & 1918 &  S   &      18  &           N  & 20.6 & 37.6 &  &   0.887  &                 &                  & \ref{J0211-8159} \\
J0255--5304     &        20   & 960 &  D   &     157  &           N  & 6.3 & 8.7 &        2  &   0.978  &   6.5 $\pm$   0.9  &   -6.3 $\pm$   0.5  & \ref{J0255-5304} \\
J0304+1932     &        20   & 959 &  D   &     298  &           N  & 11.3 & 15.8 &        2  &   0.945  &  31.5 $\pm$   0.5  &   13.0 $\pm$   0.4  & \ref{J0304+1932} \\
J0401--7608     &        10   & 5455 &  D   &     122  &           N  & 12.9 & 17.9 &        2  &   0.964  &  20.6 $\pm$   1.3  &                  & \ref{J0401-7608} \\
J0401--7608     &        20   & 7582 &  M  &     383  &           N  & 13.8 & 18.0 &        2  &   0.989  &  27.1 $\pm$   0.4  &                  & \ref{J0401-7608} \\
J0448--2749     &        20   & 960 &  S   &      58  &           N  & 9.2 & 16.7 &  &   0.992  &  19.7 $\pm$   3.6  &  -14.3 $\pm$   2.6  & \ref{J0448-2749} \\
J0450--1248     &        20   & 960 &  D   &      50  &           N  & 20.5 & 30.0 &  &   0.990  &  16.7 $\pm$   3.5  &                  & \ref{J0450-1248} \\
J0459--0210     &        20   & 958 &  D   &      38  &           N  & 3.4 & 13.2 &  &   0.890  &                 &   -9.4 $\pm$   3.1  & \ref{J0459-0210} \\
J0520--2553     &        20   & 960 &  D   &      51  &           N  & 15.1 & 19.6 &  &   0.796  &  14.0 $\pm$   3.3  &                  & \ref{J0520-2553} \\
J0525+1115     &        20   & 959 &  M  &     103  &           N  & 14.4 & 17.6 &        3  &   0.894  &   9.2 $\pm$   1.7  &    6.8 $\pm$   1.2  & \ref{J0525+1115} \\
J0533+0402     &        20   & 1918 &  S   &      81  &           N  & 5.5 & 10.1 &  &   0.996  &   7.9 $\pm$   2.4  &                  & \ref{J0533+0402} \\
J0536--7543     &        10   & 3567 &  M  &     118  &           N  & 16.6 & 25.3 &        3  &   0.918  &  29.3 $\pm$   1.4  &  -23.2 $\pm$   1.0  & \ref{J0536-7543} \\
J0536--7543     &        20   & 10775 &  M  &    1557  &           N  & 17.0 & 27.0 &        4  &   0.899  &  47.5 $\pm$   0.1  &  -11.1 $\pm$   0.1  & \ref{J0536-7543} \\
J0540--7125     &        20   & 1916 &  M  &      29  &           N  & 9.6 & 15.9 &  &   0.895  &                 &                  & \ref{J0540-7125} \\
J0543+2329     &        20   & 809 &  D   &     194  &           N  & 7.8 & 23.3 &        2  &   0.971  &  43.2 $\pm$   1.0  &   -8.9 $\pm$   0.7  & \ref{J0543+2329} \\
J0601--0527     &        20   & 960 &  D   &     138  &           N  & 16.1 & 21.8 &        2  &   0.888  &  28.8 $\pm$   1.1  &    2.9 $\pm$   0.8  & \ref{J0601-0527} \\
J0614+2229     &        10   & 180 &  S   &      47  &           N  & 8.0 & 14.6 &  &   0.983  &  46.3 $\pm$   3.8  &   16.8 $\pm$   2.8  & \ref{J0614+2229} \\
J0614+2229     &        20   & 807 &  S   &     197  &           N  & 7.4 & 13.5 &        1  &   0.997  &  69.8 $\pm$   0.9  &   18.5 $\pm$   0.6  & \ref{J0614+2229} \\
J0624--0424     &        20   & 899 &  M  &     167  &           N  & 2.0 & 21.9 &        4  &   0.736  &  21.7 $\pm$   0.8  &    6.4 $\pm$   0.6  & \ref{J0624-0424} \\
J0630--2834     &        10   & 2396 &  S   &     297  &           N  & 16.3 & 30.5 &        3  &   0.998  &  24.5 $\pm$   0.6  &   -4.2 $\pm$   0.4  & \ref{J0630-2834} \\
J0630--2834     &        20   & 5776 &  S   &     871  &           N  & 18.4 & 34.9 &        3  &   0.999  &  44.8 $\pm$   0.2  &   -3.3 $\pm$   0.1  & \ref{J0630-2834} \\
J0631+1036     &        20   & 1919 &  M  &      78  &           N  & 8.2 & 24.4 &  &   0.987  &  82.6 $\pm$   2.2  &                  & \ref{J0631+1036} \\
J0636--4549     &        20   & 1917 &  S   &      20  &           N  & 3.2 & 5.9 &  &   0.973  &                 &                  & \ref{J0636-4549} \\
J0656--2228     &        20   & 900 &  S   &      50  &           N  & 4.7 & 8.7 &  &   0.985  &                 &                  & \ref{J0656-2228} \\
J0656--5449     &        20   & 960 &  S   &      30  &           N  & 10.9 & 20.0 &  &   0.939  &                 &  -17.9 $\pm$   5.3  & \ref{J0656-5449} \\
J0659+1414     &        10   & 179 &  S   &      87  &           N  & 14.8 & 32.1 &  &   0.986  &  49.7 $\pm$   2.3  &  -14.5 $\pm$   1.7  & \ref{J0659+1414} \\
J0659+1414     &        20   & 717 &  S   &     111  &           N  & 13.9 & 30.3 &        1  &   0.984  &  70.5 $\pm$   1.7  &  -15.0 $\pm$   1.2  & \ref{J0659+1414} \\
J0709--5923     &        20   & 959 &  S   &      26  &           N  & 4.1 & 7.5 &  &   0.966  &                 &                  & \ref{J0709-5923} \\
J0719--2545     &        20   & 959 &  S   &      62  &           N  & 6.2 & 11.3 &  &   0.917  &  10.0 $\pm$   2.9  &   13.6 $\pm$   2.1  & \ref{J0719-2545} \\
J0729--1448     &        10   & 1779 &  D   &      34  &           N  & 20.0 & 28.0 &  &   0.946  &  78.9 $\pm$   5.4  &                  & \ref{J0729-1448} \\
J0729--1448     &        20   & 5146 &  D   &      94  &           N  & 10.7 & 26.7 &  &   0.886  &  72.1 $\pm$   1.9  &   16.6 $\pm$   1.2  & \ref{J0729-1448} \\
J0729--1836     &        20   & 960 &  D   &      88  &           N  & 9.8 & 18.5 &  &   0.896  &  19.6 $\pm$   2.1  &   -6.1 $\pm$   1.5  & \ref{J0729-1836} \\
J0742--2822     &        10   & 430 &  M  &     448  &           N  & 9.9 & 16.0 &        3  &   0.961  &  80.6 $\pm$   0.4  &   -4.2 $\pm$   0.3  & \ref{J0742-2822} \\
J0742--2822     &        20   & 3296 &  M  &    2463  &           N  & 9.9 & 15.9 &        3  &   0.972  &  81.4 $\pm$   0.1  &   -1.7 $\pm$   0.0  & \ref{J0742-2822} \\
J0745--5353     &        10   & 2079 &  S   &      96  &           N  & 17.7 & 32.4 &  &   0.992  &  11.9 $\pm$   1.8  &                  & \ref{J0745-5353} \\
J0745--5353     &        20   & 10012 &  S   &     476  &           N  & 18.2 & 33.6 &        2  &   0.997  &  20.8 $\pm$   0.4  &    3.6 $\pm$   0.3  & \ref{J0745-5353} \\
J0749--4247     &        20   & 959 &  S   &      44  &           N  & 4.8 & 8.8 &  &   0.963  &  17.9 $\pm$   4.1  &                  & \ref{J0749-4247} \\
J0820--3921     &        20   & 1920 &  S   &      26  &           N  & 24.2 & 44.4 &  &   0.941  &  23.2 $\pm$   6.2  &   15.5 $\pm$   4.2  & \ref{J0820-3921} \\
J0821--3824     &        20   & 1079 &  S   &      30  &           N  & 17.1 & 31.2 &  &   0.934  &                 &                  & \ref{J0821-3824} \\
J0821--4221     &        20   & 960 &  D   &      43  &           N  & 13.8 & 22.5 &  &   0.933  &                 &                  & \ref{J0821-4221} \\
J0834--4159     &        10   & 3580 &  S   &      15  &           N  & 10.8 & 20.3 &  &   0.875  &                 &                  & \ref{J0834-4159} \\
J0834--4159     &        20   & 15737 &  S   &      59  &           N  & 13.3 & 23.4 &  &   0.965  &                 &  -11.7 $\pm$   2.3  & \ref{J0834-4159} \\
J0834--4159I    &        20   & 15737 &  S   &      17  &           N  & 15.5 & 28.4 &  &   0.907  &                 &                  & \ref{J0834-4159} \\
J0835--4510     &        10   & 820 &  D   &    9275  &           N  & 6.4 & 17.2 &        4  &   0.833  &  89.6 $\pm$   0.0  &  -17.5 $\pm$   0.0  & \ref{J0835-4510} \\
J0835--4510     &        20   & 6496 &  D   &   26906  &           N  & 6.0 & 14.3 &        3  &   0.949  &  90.0 $\pm$   0.0  &   -6.0 $\pm$   0.0  & \ref{J0835-4510} \\
J0838--2621     &        20   & 1919 &  D   &      31  &           N  & 10.0 & 35.2 &  &   0.886  &                 &                  & \ref{J0838-2621} \\
J0843--5022     &        20   & 960 &  S   &      36  &           N  & 10.2 & 18.7 &  &   0.968  &                 &                  & \ref{J0843-5022} \\
J0849--6322     &        20   & 1919 &  D   &      68  &           N  & 6.5 & 22.7 &  &   0.900  &  24.8 $\pm$   2.6  &   -9.1 $\pm$   1.7  & \ref{J0849-6322} \\
J0855--4644     &        10   & 3560 &  S   &      12  &           N  & 38.0 & 70.1 &  &   0.796  &                 &                  & \ref{J0855-4644} \\
J0855--4644     &        20   & 29519 &  S   &      49  &           N  & 35.5 & 65.4 &  &   0.953  &  48.0 $\pm$   3.8  &                  & \ref{J0855-4644} \\
J0856--6137     &        20   & 959 &  S   &      94  &           N  & 7.6 & 13.8 &  &   0.997  &   8.9 $\pm$   1.9  &    5.3 $\pm$   1.4  & \ref{J0856-6137} \\
\end{tabular}
\caption{The measured profile properties of the sample of pulsars. The first column is the pulsar name (I indicates the interpulse), followed by the observing wavelength, the total integration time, the classification (Single, Double or Multiple), the signal to noise ratio of the profile, whether the profile shows some evidence for scattering (by eye), the profile width measured at the 50\% intensity point, the profile width measured at the 10\% intensity point, the number of significant fit functions to the profile for a $S/N$ of 30, the symmetry coefficient, the percentage linear polarization, the percentage circular polarization and the figure number.}
\end{table*}
\clearpage
\setcounter{table}{0}
\begin{table*}
\begin{tabular}{lcrcrcrrcccrc}
\multicolumn{1}{c}{Name} & $\lambda_\mathrm{obs}$ & \multicolumn{1}{c}{$t_\mathrm{obs}$} &  Class    &  \multicolumn{1}{c}{$S/N$} & Scat. &  \multicolumn{1}{c}{$W_{50}$}  &  \multicolumn{1}{c}{$W_{10}$}    &  $N_\mathrm{Comp}$    &  Sym.  &  \multicolumn{1}{c}{$L$}  &  \multicolumn{1}{c}{$V$}  & Fig. \\
 & [cm] & \multicolumn{1}{c}{[sec]} & & & & \multicolumn{1}{c}{[\degr]} & \multicolumn{1}{c}{[\degr]} & & & \multicolumn{1}{c}{[\%]} & \multicolumn{1}{c}{[\%]}\\
\hline
\hline
J0857--4424     &        10   & 669 &  S   &      19  &           N  & 9.6 & 17.5 &  &   0.956  &                 &                  & \ref{J0857-4424} \\
J0857--4424     &        20   & 3493 &  S   &     110  &           N  & 11.0 & 18.8 &        2  &   0.989  &                 &                  & \ref{J0857-4424} \\
J0901--4624     &        10   & 1389 &  M  &      24  &           N  & 3.1 & 18.2 &  &   0.755  &  51.1 $\pm$   6.7  &  -44.1 $\pm$   4.6  & \ref{J0901-4624} \\
J0901--4624     &        20   & 4242 &  M  &      48  &           N  & 3.2 & 26.9 &  &   0.862  &  53.3 $\pm$   3.8  &  -35.5 $\pm$   2.7  & \ref{J0901-4624} \\
J0902--6325     &        20   & 959 &  S   &     110  &           N  & 9.5 & 16.3 &        2  &   0.970  &  24.0 $\pm$   1.8  &    7.3 $\pm$   1.2  & \ref{J0902-6325} \\
J0905--4536     &        20   & 960 &  S   &      28  &           N  & 54.1 & 86.0 &  &   0.849  &                 &                  & \ref{J0905-4536} \\
\end{tabular}
\caption{continued.}
\end{table*}
\clearpage
\setcounter{table}{0}
\begin{table*}
\begin{tabular}{lcrcrcrrcccrc}
\multicolumn{1}{c}{Name} & $\lambda_\mathrm{obs}$ & \multicolumn{1}{c}{$t_\mathrm{obs}$} &  Class    &  \multicolumn{1}{c}{$S/N$} & Scat. &  \multicolumn{1}{c}{$W_{50}$}  &  \multicolumn{1}{c}{$W_{10}$}    &  $N_\mathrm{Comp}$    &  Sym.  &  \multicolumn{1}{c}{$L$}  &  \multicolumn{1}{c}{$V$}  & Fig. \\
 & [cm] & \multicolumn{1}{c}{[sec]} & & & & \multicolumn{1}{c}{[\degr]} & \multicolumn{1}{c}{[\degr]} & & & \multicolumn{1}{c}{[\%]} & \multicolumn{1}{c}{[\%]}\\
\hline
\hline
J1112--6613     &        20   & 960 &  D   &      93  &           N  & 15.3 & 22.4 &  &   0.955  &  10.4 $\pm$   1.8  &    8.5 $\pm$   1.3  & \ref{J1112-6613} \\
J1112--6926     &        20   & 959 &  D   &      64  &           N  & 11.3 & 16.4 &  &   0.970  &                 &                  & \ref{J1112-6926} \\
J1114--6100     &        20   & 959 &  S   &     141  &           N  & 11.8 & 21.0 &        2  &   0.977  &                 &                  & \ref{J1114-6100} \\
J1115--6052     &        10   & 879 &  S   &      22  &           N  & 6.7 & 12.2 &  &   0.902  &                 &                  & \ref{J1115-6052} \\
J1115--6052     &        20   & 2995 &  S   &      48  &           N  & 6.1 & 11.2 &  &   0.945  &  45.4 $\pm$   4.4  &                  & \ref{J1115-6052} \\
J1119--6127     &        10   & 1149 &  S   &      32  &           N  & 19.3 & 35.3 &  &   0.935  &  77.7 $\pm$   5.4  &  -13.0 $\pm$   3.6  & \ref{J1119-6127} \\
\end{tabular}
\caption{continued.}
\end{table*}
\clearpage
\setcounter{table}{0}
\begin{table*}
\begin{tabular}{lcrcrcrrcccrc}
\multicolumn{1}{c}{Name} & $\lambda_\mathrm{obs}$ & \multicolumn{1}{c}{$t_\mathrm{obs}$} &  Class    &  \multicolumn{1}{c}{$S/N$} & Scat. &  \multicolumn{1}{c}{$W_{50}$}  &  \multicolumn{1}{c}{$W_{10}$}    &  $N_\mathrm{Comp}$    &  Sym.  &  \multicolumn{1}{c}{$L$}  &  \multicolumn{1}{c}{$V$}  & Fig. \\
 & [cm] & \multicolumn{1}{c}{[sec]} & & & & \multicolumn{1}{c}{[\degr]} & \multicolumn{1}{c}{[\degr]} & & & \multicolumn{1}{c}{[\%]} & \multicolumn{1}{c}{[\%]}\\
\hline
\hline
J1359--6038     &        10   & 460 &  S   &     153  &           N  & 8.5 & 14.3 &        1  &   0.993  &  64.3 $\pm$   1.2  &   17.8 $\pm$   0.9  & \ref{J1359-6038} \\
J1359--6038     &        20   & 3548 &  S   &    1068  &           N  & 7.2 & 14.0 &        2  &   0.998  &  70.5 $\pm$   0.2  &   17.8 $\pm$   0.1  & \ref{J1359-6038} \\
J1403--7646     &        20   & 899 &  D   &      31  &           N  & 26.9 & 40.5 &  &   0.837  &                 &                  & \ref{J1403-7646} \\
J1406--6121     &        10   & 1420 &  S   &      45  &           N  & 3.8 & 7.1 &  &   0.995  &                 &  -65.9 $\pm$   2.3  & \ref{J1406-6121} \\
J1406--6121     &        20   & 7866 &  S   &      30  &           Y  & 15.3 & 46.0 &  &   0.877  &                 &  -42.3 $\pm$   3.8  & \ref{J1406-6121} \\
J1410--6132     &        10   & 2580 &  S   &     202  &           N  & 12.7 & 27.4 &        3  &   0.938  &  47.8 $\pm$   1.0  &                  & \ref{J1410-6132} \\
\end{tabular}
\caption{continued.}
\end{table*}
\clearpage
\setcounter{table}{0}
\begin{table*}
\begin{tabular}{lcrcrcrrcccrc}
\multicolumn{1}{c}{Name} & $\lambda_\mathrm{obs}$ & \multicolumn{1}{c}{$t_\mathrm{obs}$} &  Class    &  \multicolumn{1}{c}{$S/N$} & Scat. &  \multicolumn{1}{c}{$W_{50}$}  &  \multicolumn{1}{c}{$W_{10}$}    &  $N_\mathrm{Comp}$    &  Sym.  &  \multicolumn{1}{c}{$L$}  &  \multicolumn{1}{c}{$V$}  & Fig. \\
 & [cm] & \multicolumn{1}{c}{[sec]} & & & & \multicolumn{1}{c}{[\degr]} & \multicolumn{1}{c}{[\degr]} & & & \multicolumn{1}{c}{[\%]} & \multicolumn{1}{c}{[\%]}\\
\hline
\hline
J1549--4848     &        20   & 7715 &  S   &     100  &           N  & 9.0 & 13.7 &        1  &   0.977  &  13.0 $\pm$   1.8  &                  & \ref{J1549-4848} \\
J1549--4848I    &        20   & 7715 &  S   &      38  &           N  & 10.1 & 18.4 &  &   0.951  &  25.0 $\pm$   5.1  &                  & \ref{J1549-4848I} \\
J1551--5310     &        10   & 969 &  S   &      26  &           N  & 6.7 & 12.5 &  &   0.974  &                 &   22.7 $\pm$   6.1  & \ref{J1551-5310} \\
J1551--5310     &        20   & 4533 &  S   &      44  &           Y  & 24.8 & 75.3 &  &   0.829  &                 &   14.1 $\pm$   3.0  & \ref{J1551-5310} \\
J1557--4258     &        20   & 960 &  M  &     116  &           N  & 4.6 & 19.0 &        2  &   0.975  &  23.0 $\pm$   1.4  &  -12.3 $\pm$   0.9  & \ref{J1557-4258} \\
J1600--5044     &        10   & 460 &  D   &     331  &           N  & 6.3 & 9.2 &        3  &   0.983  &  16.5 $\pm$   0.5  &   22.6 $\pm$   0.4  & \ref{J1600-5044} \\
\end{tabular}
\caption{continued.}
\end{table*}
\clearpage
\setcounter{table}{0}
\begin{table*}
\begin{tabular}{lcrcrcrrcccrc}
\multicolumn{1}{c}{Name} & $\lambda_\mathrm{obs}$ & \multicolumn{1}{c}{$t_\mathrm{obs}$} &  Class    &  \multicolumn{1}{c}{$S/N$} & Scat. &  \multicolumn{1}{c}{$W_{50}$}  &  \multicolumn{1}{c}{$W_{10}$}    &  $N_\mathrm{Comp}$    &  Sym.  &  \multicolumn{1}{c}{$L$}  &  \multicolumn{1}{c}{$V$}  & Fig. \\
 & [cm] & \multicolumn{1}{c}{[sec]} & & & & \multicolumn{1}{c}{[\degr]} & \multicolumn{1}{c}{[\degr]} & & & \multicolumn{1}{c}{[\%]} & \multicolumn{1}{c}{[\%]}\\
\hline
\hline
J1650--4921     &        20   & 959 &  S   &      22  &           N  & 6.0 & 10.9 &  &   0.920  &                 &   19.2 $\pm$   5.8  & \ref{J1650-4921} \\
J1651--7642     &        20   & 958 &  D   &      47  &           N  & 15.9 & 22.4 &  &   0.930  &  40.9 $\pm$   4.3  &                  & \ref{J1651-7642} \\
J1652--1400     &        20   & 960 &  S   &      20  &           N  & 13.9 & 25.3 &  &   0.933  &  37.7 $\pm$   8.4  &  -25.9 $\pm$   6.1  & \ref{J1652-1400} \\
J1653--3838     &        20   & 960 &  D   &     109  &           N  & 3.7 & 14.8 &        2  &   0.860  &  29.3 $\pm$   1.5  &    4.3 $\pm$   1.1  & \ref{J1653-3838} \\
J1654--2713     &        20   & 959 &  S   &      22  &           N  & 6.7 & 12.2 &  &   0.967  &                 &                  & \ref{J1654-2713} \\
J1655--3048     &        20   & 1919 &  M  &      89  &           N  & 55.8 & 73.2 &  &   0.928  &  18.2 $\pm$   1.9  &   11.3 $\pm$   1.5  & \ref{J1655-3048} \\
\end{tabular}
\caption{continued.}
\end{table*}
\clearpage
\setcounter{table}{0}
\begin{table*}
\begin{tabular}{lcrcrcrrcccrc}
\multicolumn{1}{c}{Name} & $\lambda_\mathrm{obs}$ & \multicolumn{1}{c}{$t_\mathrm{obs}$} &  Class    &  \multicolumn{1}{c}{$S/N$} & Scat. &  \multicolumn{1}{c}{$W_{50}$}  &  \multicolumn{1}{c}{$W_{10}$}    &  $N_\mathrm{Comp}$    &  Sym.  &  \multicolumn{1}{c}{$L$}  &  \multicolumn{1}{c}{$V$}  & Fig. \\
 & [cm] & \multicolumn{1}{c}{[sec]} & & & & \multicolumn{1}{c}{[\degr]} & \multicolumn{1}{c}{[\degr]} & & & \multicolumn{1}{c}{[\%]} & \multicolumn{1}{c}{[\%]}\\
\hline
\hline
J1739--2903     &        10   & 340 &  D   &      84  &           N  & 6.5 & 9.9 &  &   0.977  &                 &                  & \ref{J1739-2903} \\
J1739--2903I    &        10   & 340 &  S   &      21  &           N  & 7.9 & 14.4 &  &   0.891  &                 &                  & \ref{J1739-2903I} \\
J1739--2903     &        20   & 2255 &  D   &     219  &           N  & 7.0 & 11.8 &        2  &   0.982  &                 &    5.5 $\pm$   0.9  & \ref{J1739-2903} \\
J1739--2903I    &        20   & 2255 &  S   &     101  &           N  & 8.0 & 15.3 &        2  &   0.975  &  19.5 $\pm$   1.8  &   -3.9 $\pm$   1.3  & \ref{J1739-2903I} \\
J1739--3023     &        10   & 1000 &  S   &      14  &           N  & 11.2 & 20.4 &  &   0.873  &  48.9 $\pm$  13.9  &                  & \ref{J1739-3023} \\
J1739--3023     &        20   & 3248 &  S   &      55  &           N  & 10.8 & 19.8 &  &   0.979  &  61.0 $\pm$   3.1  &   -9.7 $\pm$   2.2  & \ref{J1739-3023} \\
\end{tabular}
\caption{continued.}
\end{table*}
\clearpage
\setcounter{table}{0}
\begin{table*}
\begin{tabular}{lcrcrcrrcccrc}
\multicolumn{1}{c}{Name} & $\lambda_\mathrm{obs}$ & \multicolumn{1}{c}{$t_\mathrm{obs}$} &  Class    &  \multicolumn{1}{c}{$S/N$} & Scat. &  \multicolumn{1}{c}{$W_{50}$}  &  \multicolumn{1}{c}{$W_{10}$}    &  $N_\mathrm{Comp}$    &  Sym.  &  \multicolumn{1}{c}{$L$}  &  \multicolumn{1}{c}{$V$}  & Fig. \\
 & [cm] & \multicolumn{1}{c}{[sec]} & & & & \multicolumn{1}{c}{[\degr]} & \multicolumn{1}{c}{[\degr]} & & & \multicolumn{1}{c}{[\%]} & \multicolumn{1}{c}{[\%]}\\
\hline
\hline
J1828--1101I    &        10   & 570 &  S   &      23  &           N  & 7.4 & 13.6 &  &   0.955  &  52.9 $\pm$   9.2  &  -26.7 $\pm$   6.2  & \ref{J1828-1101} \\
J1828--1101     &        20   & 2738 &  S   &      57  &           Y  & 34.1 & 104.2 &  &   0.796  &  60.0 $\pm$   2.7  &   20.2 $\pm$   1.7  & \ref{J1828-1101} \\
J1828--1101I    &        20   & 2738 &  S   &      12  &           Y  & 36.6 & 67.6 &  &   0.850  &                 &  -25.5 $\pm$   6.6  & \ref{J1828-1101} \\
J1830--1059     &        10   & 689 &  M  &      60  &           N  & 2.0 & 6.7 &  &   0.986  &  65.6 $\pm$   3.3  &  -24.8 $\pm$   2.3  & \ref{J1830-1059} \\
J1830--1059     &        20   & 6333 &  S   &     156  &           N  & 2.4 & 7.1 &        3  &   0.990  &  80.1 $\pm$   1.1  &  -15.4 $\pm$   0.8  & \ref{J1830-1059} \\
J1831--0952     &        10   & 900 &  D   &      15  &           N  & 41.8 & 59.0 &  &   0.780  &  49.3 $\pm$   9.3  &                  & \ref{J1831-0952} \\
\end{tabular}
\caption{continued.}
\end{table*}
\clearpage
\setcounter{table}{0}
\begin{table*}
\begin{tabular}{lcrcrcrrcccrc}
\multicolumn{1}{c}{Name} & $\lambda_\mathrm{obs}$ & \multicolumn{1}{c}{$t_\mathrm{obs}$} &  Class    &  \multicolumn{1}{c}{$S/N$} & Scat. &  \multicolumn{1}{c}{$W_{50}$}  &  \multicolumn{1}{c}{$W_{10}$}    &  $N_\mathrm{Comp}$    &  Sym.  &  \multicolumn{1}{c}{$L$}  &  \multicolumn{1}{c}{$V$}  & Fig. \\
 & [cm] & \multicolumn{1}{c}{[sec]} & & & & \multicolumn{1}{c}{[\degr]} & \multicolumn{1}{c}{[\degr]} & & & \multicolumn{1}{c}{[\%]} & \multicolumn{1}{c}{[\%]}\\
\hline
\hline
J1901--0906     &        20   & 898 &  D   &     228  &           N  & 9.0 & 11.3 &        4  &   0.952  &  28.4 $\pm$   0.8  &                  & \ref{J1901-0906} \\
J1901--1740     &        20   & 957 &  S   &      12  &           N  & 9.7 & 17.8 &  &   0.933  &                 &                  & \ref{J1901-1740} \\
J1904+0004     &        20   & 960 &  S   &     104  &           N  & 19.1 & 34.9 &        1  &   0.983  &  23.9 $\pm$   1.6  &   14.4 $\pm$   1.2  & \ref{J1904+0004} \\
J1904--1224     &        20   & 959 &  S   &      26  &           N  & 3.2 & 11.2 &  &   0.821  &                 &                  & \ref{J1904-1224} \\
J1919+0134     &        20   & 959 &  D   &     105  &           N  & 9.6 & 17.9 &        2  &   0.927  &  19.3 $\pm$   1.6  &   -8.9 $\pm$   1.1  & \ref{J1919+0134} \\
J1932--3655     &        20   & 959 &  D   &      51  &           N  & 3.8 & 13.0 &  &   0.959  &  27.0 $\pm$   3.4  &   11.1 $\pm$   2.4  & \ref{J1932-3655} \\
\end{tabular}
\caption{continued.}
\end{table*}
\clearpage